\documentclass[10pt,leqno]{article}
\usepackage{indentfirst,csquotes}
\usepackage{graphicx}
\usepackage[hidelinks]{hyperref}
\usepackage{xurl}%

\usepackage{graphicx}%
\usepackage{multirow}%
\usepackage{amsmath,amssymb,amsfonts}%
\usepackage{amsthm}%
\usepackage{mathrsfs}%
\usepackage[title]{appendix}%
\usepackage{xcolor}%
\usepackage{textcomp}%
\usepackage{manyfoot}%
\usepackage{booktabs}%
\usepackage{algorithm}%
\usepackage{algorithmicx}%
\usepackage{algpseudocode}%
\usepackage{listings}%
\usepackage{hyperref}%
\usepackage{graphicx}%
\usepackage{natbib}
\usepackage{textcomp}

\usepackage{amsmath}

\begin{document}
\title{MedPix 2.0: A Comprehensive Multimodal Biomedical Data set for Advanced AI Applications with Retrieval Augmented Generation and Knowledge Graphs}
%
%

\author{Irene Siragusa$^{1,\ddagger}$,
Salvatore Contino$^{1,\ddagger,*}$, Massimo La Ciura$^{1}$,\\
Rosario Alicata$^{1}$,
Roberto Pirrone$^{1}$}



{\let\thefootnote\relax\footnotetext{$^{*}$corrisponding author salvatore.contino01@unipa.it (Salvatore Contino)}} 
{\let\thefootnote\relax\footnotetext{$^{\ddagger}$These authors contributes equally.}} 
{\let\thefootnote\relax\footnotetext{$^{1}$Department of Engineering, University of Palermo, 90128 Palermo, Italy} }

\maketitle              
\abstract{The increasing interest in developing Artificial Intelligence applications in the medical domain, suffers from the lack of high-quality data set, mainly due to privacy-related issues. In addition, the recent increase in Vision Language Models (VLM) leads to the need for multimodal medical data sets, where clinical reports and findings are attached to the corresponding medical scans. This paper illustrates the entire workflow for building the MedPix 2.0 data set. Starting with the well-known multimodal data set MedPix\textsuperscript{\textregistered}, mainly used by physicians, nurses, and healthcare students for Continuing Medical Education purposes, a semi-automatic pipeline was developed to extract visual and textual data followed by a manual curing procedure in which noisy samples were removed, thus creating a MongoDB database. Along with the data set, we developed a Graphical User Interface aimed at navigating efficiently the MongoDB instance and obtaining the raw data that can be easily used for training and/or fine-tuning VLMs. To enforce this point, in this work, we first recall DR-Minerva, a Retrieve Augmented Generation-based VLM model trained upon MedPix 2.0. DR-Minerva predicts the body part and the modality used to scan its input image. We also propose the extension of DR-Minerva with a Knowledge Graph that uses Llama 3.1 Instruct 8B, and leverages MedPix 2.0. The resulting architecture can be queried in a end-to-end manner, as a medical decision support system. MedPix 2.0 is available on GitHub)

%

\section{Introduction}\label{sec1}

The rise of computer-based applications in recent years has strongly favored the digitization of historically analogous processes in the management and analysis of biomedical data. In turn, the emergence of technologies based on Artificial Intelligence (AI) prompted the development of increasingly precise models to support diagnosis for the construction of personalized treatments. In the biomedical domain, in fact, different subareas can benefit from AI-based systems, ranging from the administrative field (queue management in emergency areas, centralized integration of EHRs, and so on) to the clinical one, thanks to the use of AI that can efficiently extract useful features to reliably achieve diagnosis.

One of the fundamental requirements of these applications is their trustworthiness, as they must help physicians with a high degree of confidence, being able to provide reliable predictions and classifications. However, AI models require a considerable amount of data to achieve good performance. Data availability makes expansion into the biomedical domain much more complex than in other ones. One of the main problems lies in the availability of data sets to allow the scientific community to develop new AI approaches. This problem arises mainly from the sensitive nature of data that encompass privacy issues, and this makes it difficult to build public data sets containing images and/or clinical reports to be available to the scientific community. 

To overcome these obstacles, the European Community created the European Health Data Space (EHDS)\footnote{\url{https://www.european-health-data-space.com/}}. This is a health-specific ecosystem composed of common rules, standards and practices, infrastructure, and a governance framework that aims to empower people through increased digital access and control of their personal electronic health data. The EHDS promotes a single market for electronic health record systems, relevant medical devices, and high-risk artificial intelligence systems. Finally, the EHDS aims at providing the researchers, which make use of health data with a trusted framework to control the entire analytical process \citep{penedo_2024,terzis_2023}.
The systemic approach proposed by the EHDS will provide a controlled pool of both shared data and applications. It will allow AI in the medical field to access certified and controlled data by overcoming both medical and engineering obstacles through new solid foundations on which a new generation of AI-based health applications can rely on.

In view of the implications provided by the EHDS on the application side, researchers wishing to pursue their activity in the biomedical domain by developing Vision Language Models (VLMs) should be able to find and optimize the data sets that are available in a way that maximizes the available public resources. To date, there are not so many data sets that containdical images, such as Computed Tomography (CT) and Magnetic Resonance (MR)\footnote{we well refer also to MR scans as MRI, Magneric Resonance Imaging} scans, and medical reports. One of the most well-known data sets is MedPix\textsuperscript{\textregistered}\footnote{\url{https://medpix.nlm.nih.gov/home}}, a free open-access online database of medical images, teaching cases, and clinical topics, integrating images and textual metadata provided by the National Library of Medicine (NLM), the largest collection of medical documentation in the world. This collection is part of the much more famous National Institutes of Health, which is responsible for managing health and biomedical research in the United States, actively participating in the identification of new drugs from 2010 to 2016 in collaboration with the Federal Drug Administration. Given its important contribution to science, this institution is very careful about managing the privacy issue and applies strict rules for anonymising samples stored in their repositories \cite{El_Emam_2015, Chevrier_2019}, which have also been applied to the MedPix data set used as input in the proposed work.
MedPix\textsuperscript{\textregistered} includes more than 12,000 patient case scenarios, 9,000 topics, and nearly 59,000 images. 
The contribution proposed in this paper lies in both curation and a new arrangement of MedPix\textsuperscript{\textregistered} data in a non-relational database based on MongoDB. This new arrangement of the MedPix\textsuperscript{\textregistered} structure can be regarded as a brand new data set that we called \textit{MedPix 2.0}. MedPix 2.0 makes original data accessible and well structured, as it is very easy to build views capable of creating ready-made subsets for training VLMs. Database queries have been further simplified by developing a user-friendly Graphical User Interface (GUI). To enforce usability of MedPix 2.0, the user is made able to pose her/his query, and browsing the results in very close way as in the original website.
For improved the usability of MedPix 2.0, the user can formulate their own question and browse the results in a way that is very similar to the original website. In fact, unlike the original MedPix, which was created as an educational platform, our data set allows you to dynamically build training and/or test data sets of various types (image-only, text-only or multimodal) for AI models. The use of a NoSQL database that offers a schema-on-read approach to data allows diversifying access to our data set by any downstream application, in order to simplify the querying of the data set. In particular, it is extremely simple to construct documentary data sets and prompt instructions or few shots for a downstream LLM. Lastly, since we started from a data set containing documents that were not structured but simply organised into sections, it seemed natural to use a loose structuring that reflected the organisation of the original data.

Another contribution of this work is a direct demonstration of how MedPix 2.0 makes it easy to train and test VLMs. At first, we recall the architecture and the training phase of DR-Minerva \citep{drMinerva}, a VLM developed by the authors that leverages the Minerva LLM \citep{minerva} and uses Flamingo \citep{flamingo} for multi-modality and that exploits the Retrieval Augmented Generation (RAG) approach \citep{rag}. RAG-based models are spreading in diverse domains, including the bio-medical one, reaching promising results with a retrieval module composed from both vectorstores \citep{wang2024potential,medchat} and graph structures \citep{wu2024medicalgraphragsafe}. We used our MongoDB data source to generate the train/validation/test split used for DR-Minerva.

Then, we propose a Knowledge Graph (KG) built from MedPix 2.0, using Llama 3.1 Instruct 8B \citep{llama3}.  The KG can be queried to obtain information about medical diseases and their diagnoses. The entire system, that is the KG coupled with DR-Minerva, can be queried end-to-end starting from a medical image to obtain suggestions about the most probable diagnosis  with a free-text answer, thus working as a medical decision support system.

The paper is arranged as follows: Section~\ref{sota} illustrates the State Of The Art (SOTA) for medical multimodal data sets. The details on the building process of the curated data set are reported in Section~\ref{data set} along with the implementation of the MongoDB database and the GUI. The experimental results and The architecture of DR-Minerva and the creation of the KG are reported and discussed in Section~\ref{real-world-applications}, while Section~\ref{res} illustrates the experimental results. Concluding remarks and future works are drawn in Section~\ref{future_work}. 

\section{Related Works}\label{sota}
Medical data sets for AI applications suffer from diverse problems, related both with the data and the peculiarity of the domain. First, there is a privacy issue, since clinical data contain private information about the patient, thus the process of creation of such type of data set has to start with an anonymization phase. To overcome this problem, researchers rely on either open-access and textbook data or they collaborate with hospitals to create data sets. The first two methods allow for large scalability, relying on anonymous data. On the other hand, anonymization must be carried out from scratch when dealing with hospitals. Moreover, multimodal medical data suffer from scarcity when compared with other data sets related to different domains such as MS-COCO \citep{lin2014mscoco}.

One of the most used open-access multimodal database for developing medical data sets is PubMed Central\textsuperscript{\textregistered} (PMC)\footnote{\url{https://www.ncbi.nlm.nih.gov/pmc/}}. This is a widely used free archive of biomedical scientific literature, from which it is possible to build one's own data sets via semi-automatic procedures. In PMC, data are anonymized, and high-quality captions can be extracted from the medical research papers the images belong to. The following multimodal data sets were extracted from PMC: ROCO \citep{pelka2018radiology}, MedICaT \citep{subramanian2020medicat}, PMC-OA \citep{lin2023pmc}. ROCO contains pairs of radiology images and the corresponding captions, and it incorporates an out-of-class set to improve prediction and classification performances. MedICaT is a disjoint data set from ROCO that is mainly composed by radiology images and provides manually annotations for sub-figures. PMC-OA is the larger than the previous ones, and it keeps a variety of diagnostic procedures, diseases, and findings, while introducing sub-figure separation. 

VQA-RAD \citep{lau2018dataset} is a data set derived from MedPix\textsuperscript{\textregistered}, and it collects a subset of radiological images, while providing Question-Answer (QA) pairs validated by domain experts.

Another source of available high-quality data are textbooks: PathVQA \citep{he2020pathvqa} is a Visual Question Answering (VQA) data set that collects both closed- and open-ended QA pairs, which are extracted from both pathology textbooks and online digital libraries via a semi-automated pipeline.

On the other side, open access data sets like MIMIC-CXR \citep{johnson2019mimic}, IU-Xray
\citep{demner2016preparing} and SLAKE \citep{liu2021slake} are manually annotated by domain experts. Both MINIC-CXR and IU-Xray are chest radiography data sets derived from hospital's clinical cases. Both if them contain a semi-structured radiology report, describing the radiological findings of the images it is related to. SLAKE, on the contrary, collects images from different radiology open-access data sets and provides manual annotations and QA pairs given by experienced doctors in English and Chinese. 

INSPECT \citep{huang2023inspect} is a worth mentioning multimodal data set that collects computed tomography pulmonary angiography (CTPA), radiology reports and structured data from EHRS, for pulmonary embolism diagnosis.

The aforementioned data sets are multimodal ones, and they are deeply focused on VQA tasks. Also some specialized data sets exist that provide only visual data, as UniToChest \citep{chaudhry2022unitochest}, ADNI \citep{adni}, OMI-DB \citep{OPTIMAM} or OASIS-3 \citep{oasis3} suitable for image segmentation and classification, or free-form textual data such as in Named Entities in Medical Case Reports (NE-MCR) in which the main focus is towards Named Entity Recognition (NER) \citep{schulz-etal-2020-named}. Other works worth mentioning are UniToBrain \citep{unitobrain} that integrates a technical report with the scanning modality, and the E3C corpus \citep{Magnini2020TheEP} that is an annotated multilingual data set of clinical reports suitable for NER and Relation Extraction (RE) tasks.

Despite the proposed MedPix 2.0 data set may be close to VQA-RAD, they differ from both the sampling strategy and the samples themselves. In Med Pix 2.0, we do not integrate QA pairs, but different QA pairs can be created relying on the structured textual information provided in the data set itself, following the structure of the other one. Thus, the creation of more complex tasks is possible, such as document summarization or understanding. 

With respect to the existing multimodal data sets, MedPix 2.0:
\begin{enumerate}
    \item is derived from a freely open-access source, and it has no privacy-related issues;
    \item offers a balanced variety of CT and MRI scans of different body parts;
    \item for each image, a complete structured clinical case is provided.
\end{enumerate}
Thanks to the annotation scheme we selected for the JSON documents in MedPix 2.0, the last point makes it suited for various tasks not only limited to document-level retrieval. Unfortunately, images are provided in PNG format thus limiting visual processing with respect to raw images in DICOM format. A comprehensive overview of the reported data set, is reported in Table \ref{tab:tax}.

\begin{table}[!h]
    \centering
    \resizebox{\textwidth}{!}{\begin{tabular}{c|c|c|c|c}
        \toprule
        \textbf{Data set}&\textbf{Reference}&\textbf{\# samples}&\textbf{Data source}&\textbf{Task}\\
        \midrule
        \multicolumn{5}{c}{\textit{Textual-only}}\\
        \midrule
        NE-MCR &\citep{schulz-etal-2020-named}&53&PMC&NER\\
        E3C&\citep{Magnini2020TheEP}&10.034&PMC&NER, RE\\
        \midrule
        \multicolumn{5}{c}{\textit{Visual-only}}\\
        \midrule
        UniToBrain&\citep{unitobrain}&258&Hospital&Image segmentation\\
        UniToChest&\citep{chaudhry2022unitochest}&306.440&Hospital&Image segmentation\\
        OASIS-3&\citep{oasis3}&6.471&Hospital&Image segmentation\\
        OMI-DB&\citep{OPTIMAM}&3.072.878&Hospital&Image classification\\
        ADNI&\citep{adni}&1921&Hospital&Image classification\\
        \midrule
        \multicolumn{5}{c}{\textit{Visual and textual}}\\
        \midrule
        ROCO&\citep{pelka2018radiology}&87.952&PMC&Multimodal classification\\
        MedICaT&\citep{subramanian2020medicat}&217.060&PMC&subfigure-subcaption alignment\\
        PMC-OA&\citep{lin2023pmc}&1.650.00&PMC&Multimodal classification\\
        VQA-RAD&\citep{lau2018dataset}&3.515&MedPix\textsuperscript{\textregistered}&VQA\\
        MINIC-CXR&\citep{johnson2019mimic}&377.110&Hospital&Classification, text generation\\
        IU-Xray&\citep{demner2016preparing}&8.121&Hospital&Multimodal retrieve\\
        SLAKE&\citep{liu2021slake}&14.028&Open data&VQA\\
        PathQA&\citep{he2020pathvqa}&32.799&Pathology books&VQA\\
        INSPECT&\citep{huang2023inspect}&23.248&Hospital&Multimodal classification\\
        
        \midrule
        \textbf{MedPix 2.0}&&2050&MedPix\textsuperscript{\textregistered}&Multimodal classification, text generation\\
         \bottomrule
    \end{tabular}}
    \caption{Data set taxonomy}
    \label{tab:tax}
\end{table}

\section{MedPix 2.0}\label{data set}

MedPix\textsuperscript{\textregistered} is a free open-access multimodal online database of medical images, teaching cases, and clinical topics, managed by the National Library of Medicine (NLM) of the National Institutes of Health (NIH). It mainly serves as a support system for Continuing Medical Education (CME) of physicians, nurses, and healthcare students. The database collects clinical cases related to more than 12,000 patients. Each case contains at least one medical image, and the corresponding findings, discussion notes, diagnosis, differential diagnosis, treatment, and follow up. Textual information is reported in a semi-structured format. Attached to the clinical case, there is the topic section, where the disease under investigation is discussed in detail from an academic and general perspective. 

In Fig.~\ref{medpixoverview} an example from the MedPix\textsuperscript{\textregistered} website is reported.

\begin{figure}
\includegraphics[width=\textwidth]{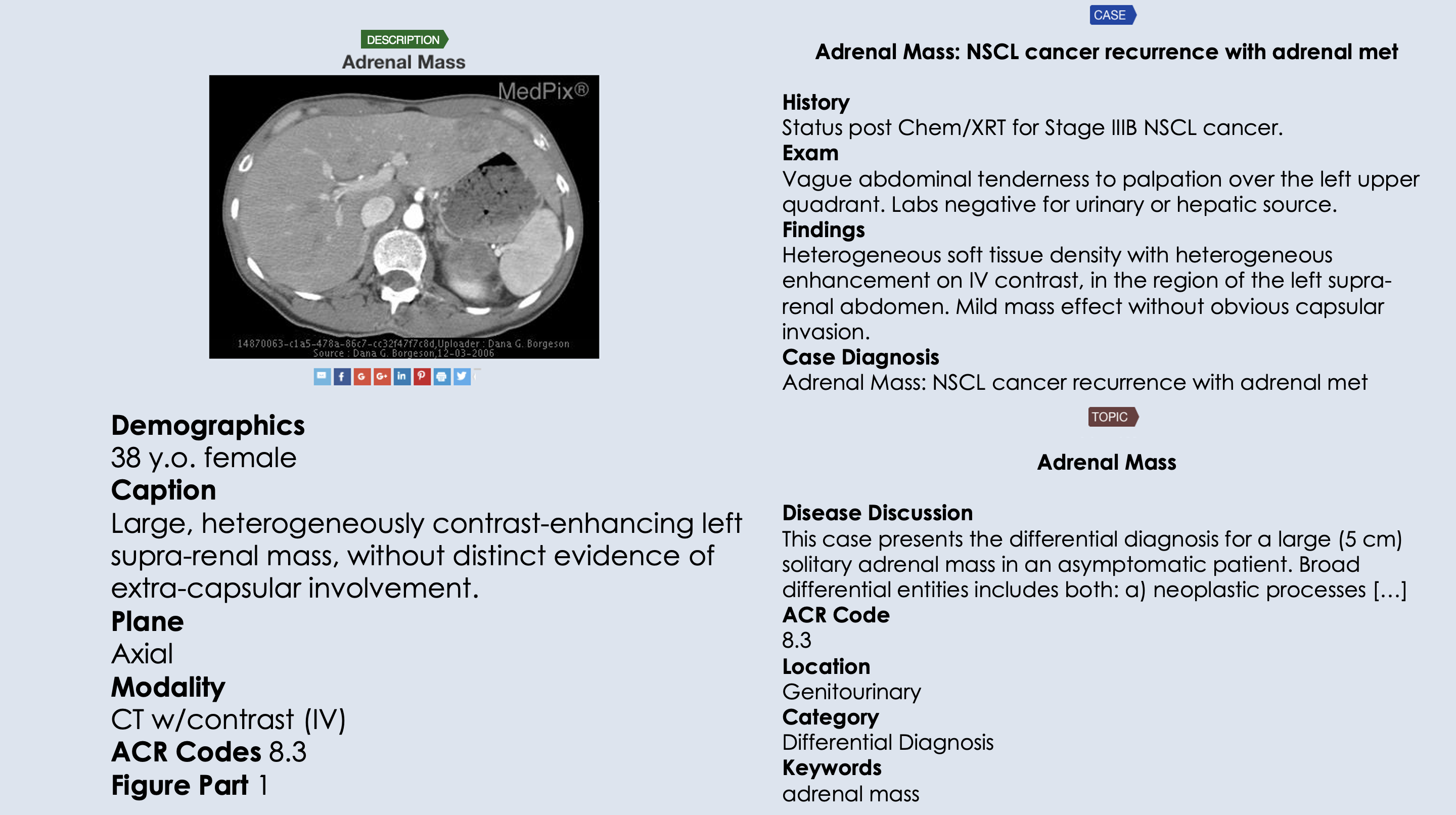}
\caption{An example from the MedPix\textsuperscript{\textregistered} web site where the original screenshots have been rearranged for visualizing purposes, and the colored labels have been added for clarity.} \label{medpixoverview}
\end{figure}

Despite the richness of the data set, its free availability, the possibility to add new cases, and the access to clinical cases of interest using wither keywords, the body part or the disease, it is not possible to access the raw data. This feature limits the usage of MedPix\textsuperscript{\textregistered} for training multimodal AI systems. Therefore we decided to create a brand new structured version of this data set because it represents a high quality source for AI-based medical applications. MedPix 2.0 has been built essentially as a MongoDB instance that is released along with a suitable GUI aimed both at general purpose querying and extracting training data for AI models. Referring to the Fig.\ref{medpixoverview}, a MongoDB version of the data set was built using a semi-automated pipeline to create two kinds of JSON documents: the one collecting the information falling into the screenshots labeled as \emph{DESCRIPTION}, and the one that gathers the information falling into the screenshots labeled both as \emph{CASE} and \emph{TOPIC}. As mentioned in the section \ref{sec1}, the choice of MongoDB is motivated by the aim of respecting the organisation of the main data, improving its readability. In fact, the pairs of images and texts are connected to each other by reference URIs, making it unnecessary to store the images within the documents as is typically done with a relational approach. In this way the creation of multimodal data sets is made more intuitive, facilitating any downstream application.

\subsection{Data set extraction}

\begin{figure}[h!]
    \centering
    \includegraphics[width=\textwidth]{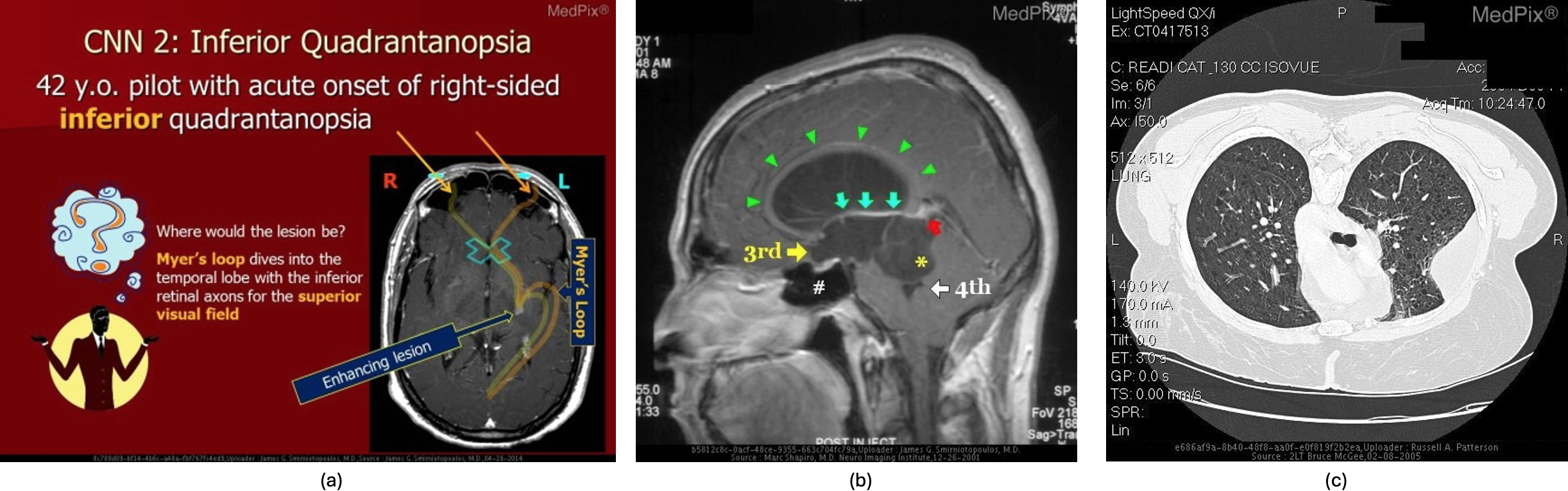}
    \caption{Among the manually removed images there are presentation's slide (a), manually (b) and automatic annotated images from DICOM viewer (c)} 
    \label{noisy_samples}
\end{figure}

We decided to focus on a part of MedPix\textsuperscript{\textregistered} that involves cases related with two diagnostic modalities, namely Computed Tomography (CT), and Magnetic Resonance Imaging (MRI). First of all, the images in the considered split were downloaded via Open-i\textsuperscript{\textregistered} \footnote{\url{https://openi.nlm.nih.gov/s}}: among them, there where some noisy images such as teaching materials or annotated images (Fig. \ref{noisy_samples}) that are inconsistent as input image for training a VLM for which we expect a clear CT or MRI.
In particular, this process of curing the data set by removing noisy samples, was performed via manual direct inspection of all the automatically downloaded samples. This was the only non-automated task performed during the data set extraction process; therefore, we refer to our approach as semi-automated. This selection was crucial, because the use of images with artefacts, such as the ones shown in the figure (e.g. symbols, annotations), would not allow a correct visual feature extraction, leading to relevant biases in the classification process of the multimodal model.

A subsequent automatic scraping pipeline was implemented to extract the textual data related to the selected images using Selenium\footnote{\url{https://www.selenium.dev/}} and Beautiful Soup\footnote{\url{https://www.crummy.com/software/BeautifulSoup/bs4/doc/}}. Finally, two kinds of JSON documents were devised to store, respectively, the information strictly connected with the images (\texttt{descriptions} document) and the one related to a clinical case (\texttt{case-topic} document). A one-to-many relation has been created between a clinical cases and images by embedding the \texttt{U\_id} defined for a \texttt{case-topic} document in each \texttt{descriptions} document attached to each image related to the clinical case itself. An example of the two kinds of JSON documents is shown in 
Fig \ref{json_raw_medpix-desc} and \ref{json_raw_medpix-case}.

\begin{figure}[!h]
    \centering
    \begin{verbatim}
        
        {
            "Type": "CT",
            "U_id": "MPX1009",
            "image": "MPX1009_synpic46283",
            "Description": {
                "Caption": "The prostate is enlarged with several
                calcifications noted within. No dominant prostate mass
                is evident.",
                "Plane": "Coronal",
                "Modality": "CT - noncontrast",
                "ACR Codes": "8.-1",
                "Sex": "male",
                "Age": "73"
            }
        }
        
    \end{verbatim}
    \caption{Example of a JSON document in MedPix 2.0 as for \textit{description document}.}
    \label{json_raw_medpix-desc}
\end{figure}

\begin{figure}[!h]
    \centering
    \begin{verbatim}
        
        {
            "U_id": "MPX1009",
            "TAC": ["MPX1009_synpic46283", "MPX1009_synpic46295"],
            "MRI": [],
            "Case": {
                "Title": "Bladder Diverticulum",
                "History": "73-year-old male with hematuria and
                numerous white blood cells found on UA",
                "Exam": "N/A",
                "Findings": "Bladder with thickened wall and
                diverticulum on the right. Diverticulum is mostly
                likely secondary to chronic outflow obstruction.
                Prostate enlargement.",
                "Differential Diagnosis": "Bladder Diverticulum",
                "Case Diagnosis": "Bladder Diverticulum",
                "Diagnosis By": "N/A"
            },
            "Topic": {
                "Title": "Bladder Diverticulum",
                "Disease Discussion": "Bladder diverticula most
                often occur as a result of outlet obstruction.
                Occasionally, a congenital weakness in the bladder
                wall adjacent to the ureteral orifice results in a
                diverticulum. This is termed a "Hutch" diverticulum.
                In children, outlet obstruction causing a diverticulum
                is rare and can be seen with urethral [...]	.",
                "ACR Code": "8.9",
                "Category": "Diverticulum",
           }
        }
        
    \end{verbatim}
        \caption{Example of a JSON document in MedPix 2.0 as for \textit{case-topic document}.}
        \label{json_raw_medpix-case}
\end{figure}

\subsection{MongoDB representation}\label{mongodb}
In order to process properly the data in MedPix 2.0 we built a MongoDB database to host all the JSON documents along with the images. The architectural choice stems not only from the nature of the data in MedPix 2.0 but also from considerations related to its high flexibility and scalability to distributed environments where also private multimodal medical data could be added to the original collections with the constraint of not being moved away from their generation site as it is the case of hospital generated information.

We built a MongoDB instance made by two collections, namely \texttt{Image\_} \texttt{Descriptions}, containing the \texttt{descriptions} documents, and \texttt{Clinical\_}\texttt{re\-ports} which contains all the \texttt{case-topic} documents. In our implementation, the images are stored in a separated folder, and they are accessed using a proper \texttt{file://} URL built starting from their \texttt{U\_id}. Finally, a view called \texttt{Image\_} \texttt{Reports} allow a direct access to both collections via their \texttt{U\_id}.

A user-friendly GUI was built using PyQt5\footnote{\url{https://www.riverbankcomputing.com/software/pyqt/}} to allow an easy access to the database, querying it to obtain the desired data for visualization and/or download. As shown in Fig.~\ref{query}.a, it is possible to select either the collection or the view to be queried and, add the query input in the relative fields. In Fig.~\ref{query}.b, an example of the answer to a query is reported: samples matching the query are reported as a list of JSON objects and the user can save it or view a specific clinical case and/or image selected in list of the query results. An example of this view is shown in Fig.~\ref{interface}, where a MedPix\textsuperscript{\textregistered} like GUI is reproduced to enforce usability for the users of the original website. In our GUI, the curated images scraped along with the texts are showed by default, but the user can choose to download the original (non curated) data related to the clinical case under investigation.

\begin{figure}[!h]
\includegraphics[width=\textwidth]{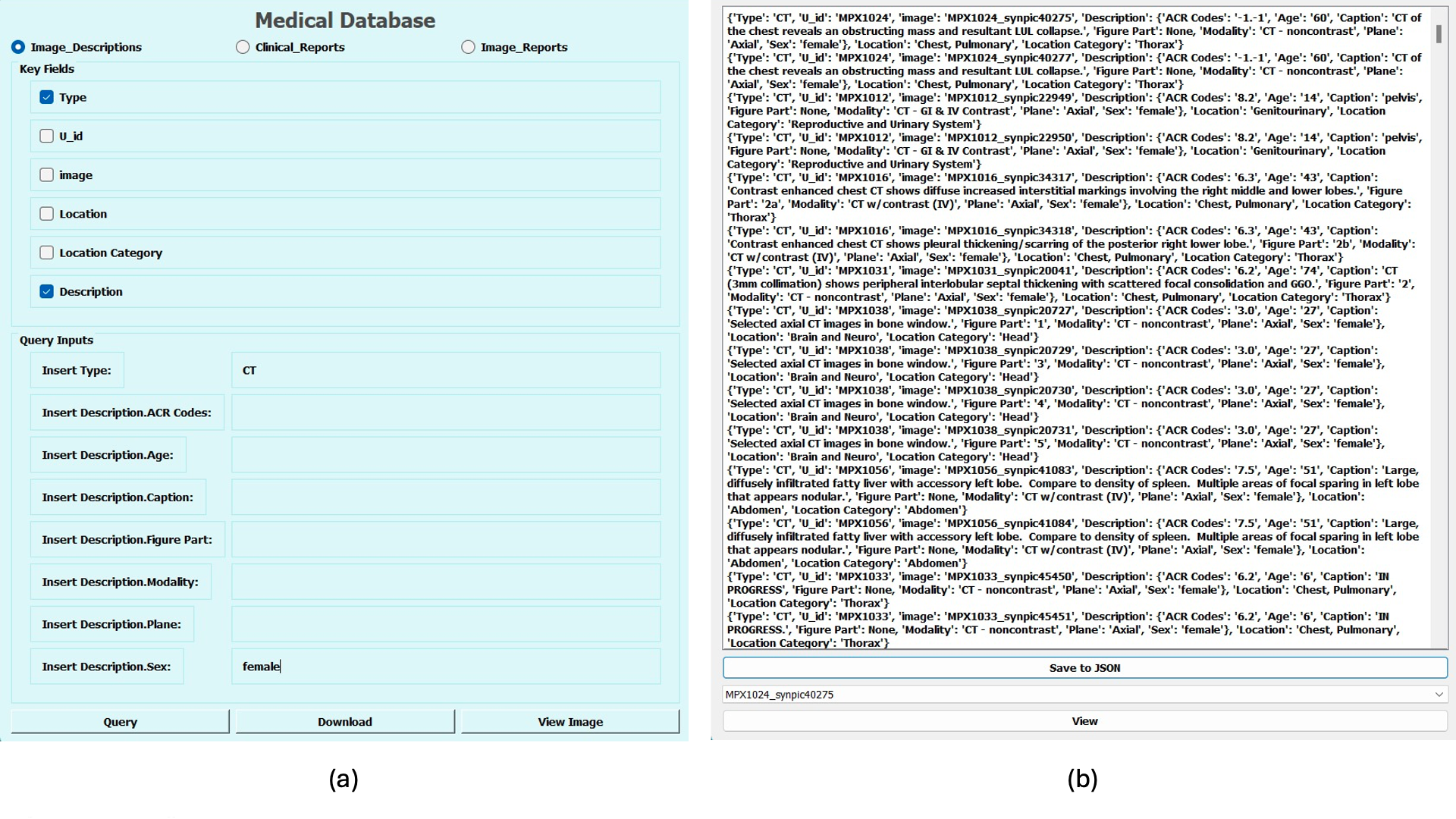}
\caption{An example of the interface for querying the database where the user asks for CT scans of female patients (a) and the provided output (b).}
\label{query}
\end{figure}

\begin{figure}[!h]
\includegraphics[width=\textwidth]{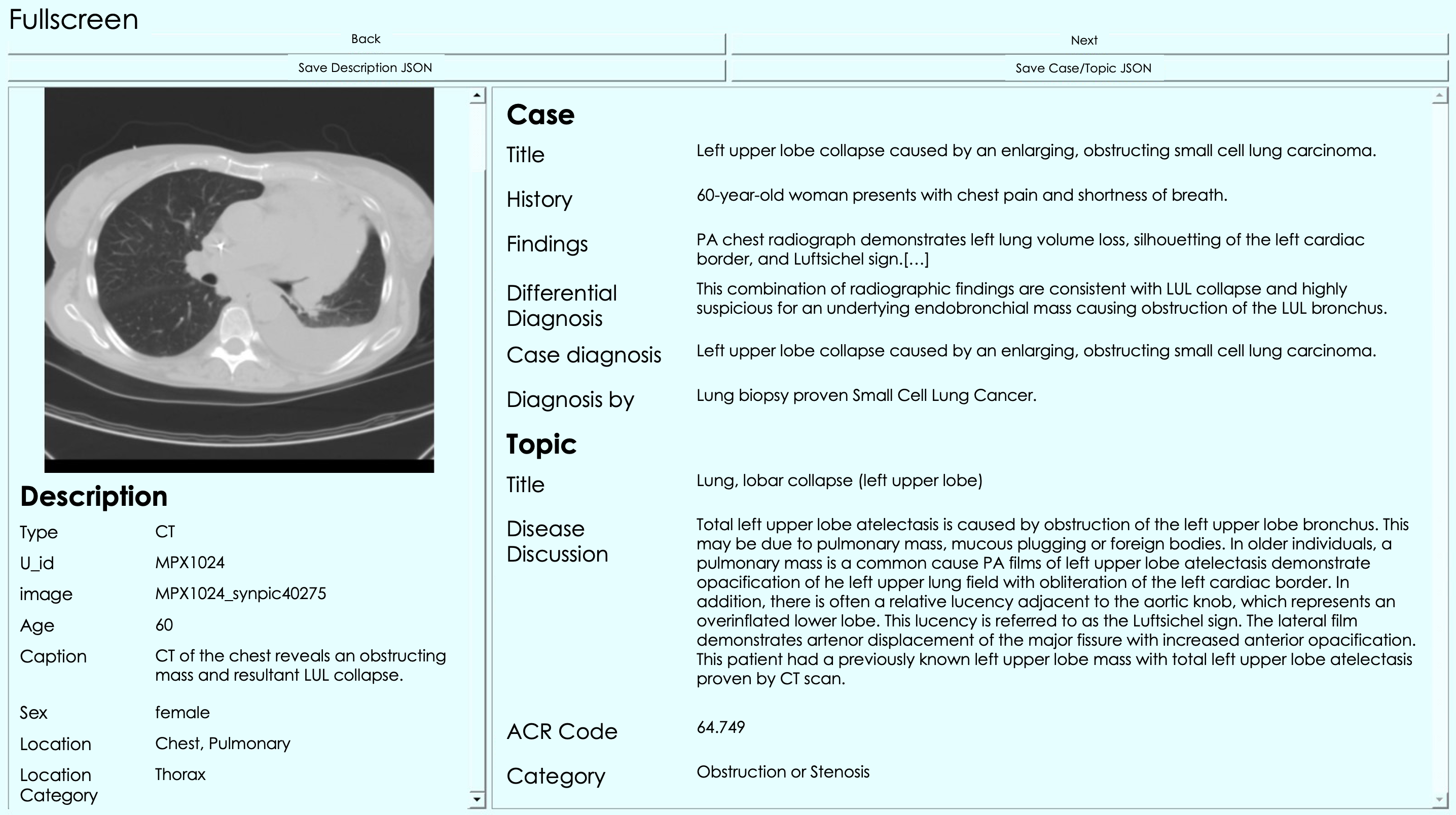}
\caption{An example of the view of a selected clinical case retrieved by the query posed in Fig. \ref{query}. The image, and the relative description are reported along with the information of the clinical case the image belongs to.} \label{interface}
\end{figure}

MedPix 2.0 and its querying interface can be a valid tool for both physicians and AI researchers since the desired data can be easily downloaded for further application, like training Deep Learning (DL) models where a large amount of structured data are required. The structured output downloaded from the interface, can be used in turn to fill a fixed template, and to provide a VLM with a rich textual prompt. Also multimodal tasks can be addressed if the textual descriptions are combined with the corresponding image, as we will demonstrate in Section~\ref{real-world-applications}. Our re-arranged version of MedPix\textsuperscript{\textregistered}, contains the same information of the original data set, but allows a easier usage for DL application in text-only, image-only and multimodal setup. This flexible structure is also suitable for adding new samples that meet the characteristics of the data set, i.e. they should be a clear CT or MR scan with the associated textual data as in Figures \ref{json_raw_medpix-desc} and \ref{json_raw_medpix-case}. The interface can be locally executed as reported in \url{https://github.com/CHILab1/MedPix-2.0/tree/main/MongoDB-UI}.

\section{Training a VLM using MedPix 2.0}\label{real-world-applications}
This section reports an application scenario for using Medpix 2.0. Specifically, a training pipeline for a multimodal model was implemented and, starting from the images, through a RAG-based approach, textual and visual embedding are generated to predict the scan modality and the body part shown in the input image. The output at this stage is furthermore used to query a generative Large Language Model (LLM) which will generate a consistent answer for the end user through the use of a Knowledge Graph (KG), the answer contains a diagnostic suggestion, leveraging the acquired knowledge on both scan modality and body part and the knowledge in the KG.

\subsection{Data preparation}

The training phase of the model was conducted using MedPix 2.0, carefully divided into three splits: Train, Validation, and Test. In order to reduce data leakage, when splitting the data set into Train, Validation and Test splits, all the images referring to the same clinical case are reported in the same split. In addition, we ensured a correct stratification of the data as for balancing both the typology of the scanning modality, CT and MR scans, and the body part shown. Body parts were grouped into five macro categories, namely Abdomen, Head, Reproductive and Urinary System (RUS), Thorax and Spine and Muscles (SaM). Splits were created using a 80:20 in which Training and Testing split were derived; from a subsequent 50:50 split over the testing split, Validation and Test splits were expunged.

The data obtained from this division are shown in Table \ref{tab:tab_dataset} and \ref{tab:tab_parti_corpo}.

\begin{table}[!h]
    \centering
    \begin{tabular}{c|c|c|c}
    \toprule
    Data set&\#CT&\#MR&Clinical Case\\
    \midrule
    Train&878&775&535\\
    Validation&84&113&67\\
    Test&100&100&69\\
    \midrule
    Total&1062&988&671\\
    \bottomrule
    \end{tabular}
    \caption{Summary of images and clinical cases for each split.}
    \label{tab:tab_dataset}
\end{table}

\begin{table}[!h]
    \centering
    \begin{tabular}{c|c|c|c|c|c}
    \toprule
    Data set&Abdomen&Head&RUS&Thorax&SaM\\
    \midrule
    Train&264&742&127&263&257\\
    Validation&23&66&20&30&58\\
    Test&32&76&11&41&40\\
    \midrule
    Total&319&884&158&334&355\\
    \bottomrule
    \end{tabular}
    \caption{Summary of images divided per macro area of the body for each split. 
    }
    \label{tab:tab_parti_corpo}
\end{table}

\subsection{RAG-based Flamingo}
Figure \ref{fig:fig_flamingo} illustrates the overall RAG architecture used to test MedPix 2.0, which is called DR-Minerva \citep{drMinerva}. DR-Minerva relies on Flamingo \citep{flamingo} and Minerva \citep{minerva}, a new LLM trained from scratch on English and Italian data as part of the activities in the PNRR FAIR Transversal Project 2: ``Vision, Language, and Multimodal Challenges''\footnote{\url{https://fondazione-fair.it/en/transversal-projects/tp2-vision-language-and-multimodal-challenges/}}. FAIR TP2 is a collaborative research carried out by more than twenty Italian universities. The authors are engaged in developing VLM for the biomedical domain. We chose Minerva because it is totally open. Not only the weights, but also the architecture is free along with the training data. Furthermore, its training set is half English and half Italian, making it suitable for future experimentation using Italian data directly.

\begin{figure}[!h]
    \centering
    \includegraphics[width=0.8\linewidth]{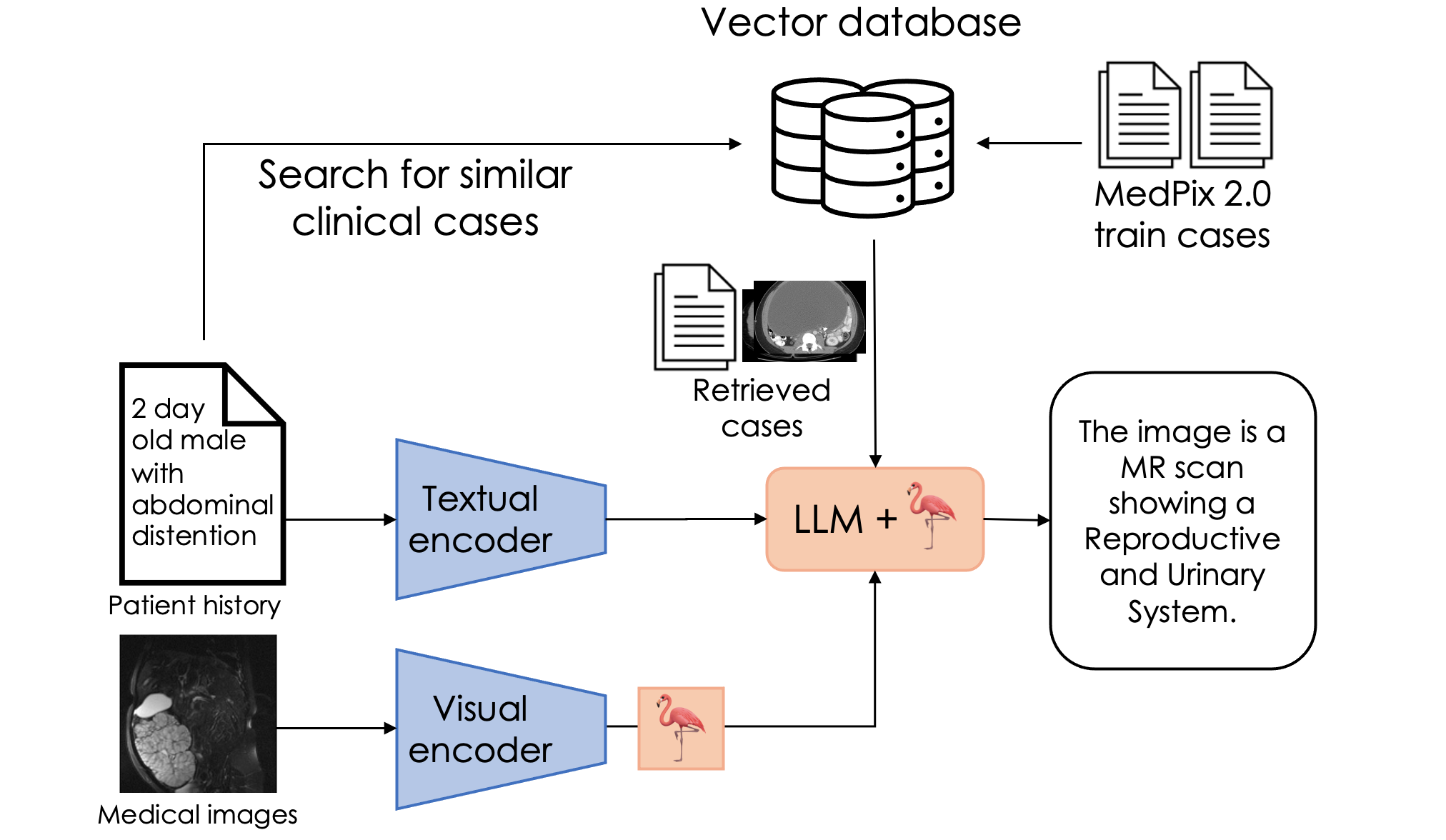}
    \caption{Overview of DR-Minerva architecture.}
    \label{fig:fig_flamingo}
\end{figure}

Flamingo was chosen as a suitable VLM since its strong in-context learning capabilities that make it suitable to various domains \citep{flamingo}. Both in training and inference phase, visual tokens are incorporated with textual encodings, via a suitable trained cross-attention layer. More in detail, Flamingo allows the flexibility to adopt a custom  visual and textual encoder, thus SOTA models can be easily used in this hybrid pipeline. On this basis, this approach was preferred to other VLMs, such as CLIP \citep{clip}, which leverages on encoder-only models as for textual encoding, thus resulting in poor generative capabilities with respect to decoder-only models. In addition, the proposed classification via CLIP, is limited to pre-defined categories, which prevent the generative capabilities of the model in providing verbose answers.
In our implementation, we used Open Flamingo \citep{awadalla2023openflamingo} in the 3B parameter version\footnote{\url{https://huggingface.co/openflamingo/OpenFlamingo-3B-vitl-mpt1b}}, where CLIP ViT-L/14 \citep{clip} was employed as visual encoder and Minerva-3B \footnote{\url{https://huggingface.co/sapienzanlp/Minerva-3B-base-v1.0}} as language encoder.
The table \ref{tab_rag_flamingo} shows the prompt used to query the model in order to obtain the desired response, which consists of both the scanning modality and the body part shown in the input image.

We developed a suitable RAG component for DR-Minerva \citep{rag} to improve the overall performance of the model at inference time, due to a prompt enriched with relevant information for the query.
The RAG technique allows you to construct a prompt that is based not only on the instructions and input provided, but also on relevant \textit{ad hoc} information retrieved from a suitable Knowledge Base and is used as state of the art to respond to tasks of this type (\cite{Jiang_Fang_2024, Salemi_Zamani_2024, Lee_2024}). On this basis, the selected LLM can produce a coherent answer, thus mitigating any hallucinations due to the intrinsic knowledge in the LLM \citep{rag}. In addition, updated knowledge can be injected to the model via an in-context strategy \citep{gpt3}, without requiring an additional fine-tuning of the model after a knowledge update \citep{peng2023checkfactstryagain}. Various strategy have been proposed to effectively enhance retriever capabilities, leveraging vector databases \citep{vidivelli2024efficiency,wang2024potential}. Obviously, increasing retrieve performances, lead to better generated outputs and in optimizing computational costs.

Because AI models must be as precise as possible, especially in the medical sector, we queried DR-Minerva with both the target medical image and a template made up of some personal information about the patient (e.g. age and gender), followed by the patient's history. The most relevant clinical cases are then retrieved and added to the prompt in the form of few-shot learning examples, based on the patient's history. The RAG component selects the k-closest clinical cases to the query, more specifically, in our experiments k is equal to four. The value of this threshold was empirically found since it resulted to the the optimal trade-off as both for providing a sufficient context to the model and as for reaching satisfactory results.
The vector database in the RAG component has been filled suitably with textual information from clinical cases, which belong to the train split, and involve the age and gender of the patient, the relative clinical history, and the doctor's diagnosis.

Our RAG is built using the LangChain framework\footnote{\url{https://www.langchain.com/}}, which in turn employs a FAISS vector database \citep{faiss}, where the data are saved using Linq-Embed-Mistral \citep{embedMistral}. This model is regarded as the best in the Massive Text Embedding Benchmark (MTEB) \citep{muennighoff2022mteb} for Information Retrieval\footnote{as in \url{https://huggingface.co/spaces/mteb/leaderboard} in June 2024}.

Both text-encoding and text generation are demanded to Minerva LLM, a transformer decoder-only LLM (\citealp{attention2017}) built upon the Mistral architecture (\citealp{jiang2023mistral7b}) which was trained from scratch on 660B tokens, equally balanced between English and Italian. Flamingo leverages a perceiver-resampler network to extract visual tokens from the visual encoding. In turn, visual tokens are employed to condition the textual output generation by inserting cross-attention layers between the existing pretrained ones and frozen LLM layers \citep{flamingo}.

\begin{table}
\resizebox{\textwidth}{!}{\begin{tabular}{l|l}
    \toprule
    \textbf{Prompt}&\textbf{Response}\\
    \midrule
    \begin{tabular}{@{}l@{}}
        Given the following medical images and the patient\\
        history, provide information about the scanning\\
        modality and the body part shown in the image.\\
        \(\bigl \langle \text{image} \bigr \rangle \) \(\bigl \langle \text{age} \bigr \rangle \) \(\bigl \langle \text{sex}  \bigr \rangle \) patient.\\
        \(\bigl \langle \text{clinical history of the patient} \bigr \rangle \).\\
        \lbrack ... context ... \rbrack\\
        The image is a \(\bigl \langle \text{scan type} \bigr \rangle \) scan showing a \(\bigl \langle \text{body part} \bigr \rangle \).\\
        \lbrack ... context ...\rbrack \\
        \(\bigl \langle \text{image}\bigr \rangle \) \(\bigl \langle \text{age} \bigr \rangle \) \(\bigl \langle \text{sex} \bigr \rangle \) patient.\\
        \(\bigl \langle \text{clinical history of the patient} \bigr \rangle \).
    \end{tabular} &
    \begin{tabular}{@{}l@{}}
        The image is a\\
        \(\bigl \langle \text{predicted scan type} \bigr \rangle \)\\
        scan showing a\\
        \(\bigl \langle \text{predicted body part} \bigr \rangle \).
    \end{tabular} \\
    \bottomrule
    \end{tabular}}
    \caption{The structure of the prompt is reported as well as the template of the corresponding expected answers}
    \label{tab_rag_flamingo}
\end{table}

\subsection{Knowledge Graph definition}\label{kg_definition}
MedPix 2.0 is actually a huge archive of historical clinical cases that can be exploited as external knowledge for sophisticated textual predictions. In addition to the information related to the particular case reported in the \textit{Case} section, such as \textit{Findings} and \textit{Discussion}, there is also more general information related to the disease, like \textit{Disease Discussion} in the \textit{Topic} section (Figures \ref{json_raw_medpix-desc} and \ref{json_raw_medpix-case})). Due to the inherent domain-specific semantics of this kind of textual information, classical RAG relying on vector databases is not efficient for precise retrieval. As a consequence, we decided to build a KG, which relies on both case-specific and general medical information that can be retrieved at inference time to generate a diagnostic suggestion for the provided input. The whole process for creating the KG and inferring suggestions is depicted in Fig. \ref{fig:kg-creation-inference}

\begin{figure}[!h]
    \centering
    \includegraphics[width=0.8\linewidth]{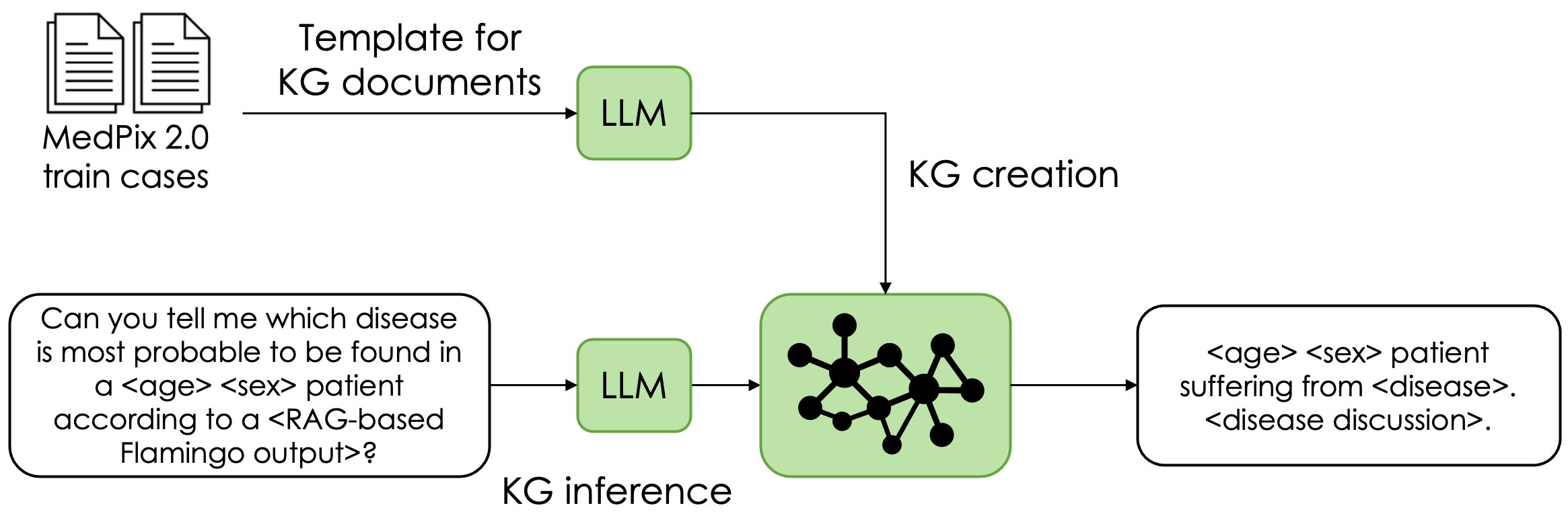}
    \caption{Overview of the overall process for KG creation and inference.}
    \label{fig:kg-creation-inference}
\end{figure}

To allow efficient relation extraction for building the KG, clinical cases belonging to the training set were arranged as documents according to the following template:

\begin{quote}
\noindent\fbox{%
    \parbox{0.92\textwidth}{%
    \(\bigl \langle \text{U\_id} \bigr \rangle \) is a clinical report of a \(\bigl \langle \text{age} \bigr \rangle \) \(\bigl \langle \text{sex} \bigr \rangle \) patient suffering from a \(\bigl \langle \text{disease} \bigr \rangle \)\ displayed in \(\bigl \langle \text{scan modality} \bigr \rangle \).\\
    \( \bigl \langle \text{clinical history of the patient} \bigr \rangle \)\\
    The disease \(\bigl \langle \text{disease name} \bigr \rangle \) located in \(\bigl \langle \text{specific body part} \bigr \rangle \) (\(\bigl \langle \text{body part} \bigr \rangle \)).\\
    \( \bigl \langle \text{clinical history of the patient} \bigr \rangle \)\\
    \(\bigl \langle \text{Treatment and followup recommended from the doctor} \bigr \rangle \).\\
    About \(\bigl \langle \text{disease} \bigr \rangle \) we can say that: \(\bigl \langle \text{disease discussion} \bigr \rangle \).
    }%
}
\end{quote}

We used LlamaIndex \citep{Liu_LlamaIndex_2022} to generate the KG. In particular, the KG is built starting from document chunks whose fixed size is 4096 tokens. At most \textit{n} triplets were extracted from each chunk to form the nodes and edges of the KG. Triplets were extracted using the default LlamaIndex query posed to Llama 3.1 8B Instruct \citep{llama3}. 


Llama 3.1 8B Instruct was chosen since it is an open-source and multi-lingual model trained with instruction-tuning strategy. More in detail, instruction tuned models, have the ability to provide a coherent output given a well-defined instruction, optionally enriched via some examples \citep{gpt3, agmodels}. This peculiar characteristics, made it suitable for both KG creation and the subsequent answer generation: both tasks rely on the capabilities of an external LLM to output a coherent answer leveraging the instruction. In addition, no further costly fine-tuning is requested at this step. No other models were considered at this step, since, at the time of developing the model: this was the latest multilingual instruct model available, trained with a Supervised Fine-Tuning Strategy with Direct Preference Optimization \citep{dpo} for human preference alignment \citep{llama3}.

We created six different KGs according to the maximum number of triplets to be extracted from each chunk (3, 5, 10) and the option to process the chunk in lowercase or not. The hyperparameter setups are reported in Table \ref{tab:kg-setup}.

\begin{table}
    \begin{tabular}{c|c|c|c|c}
        \toprule
        \textbf{Setup name}&\textbf{max rel. per chunk}&\textbf{Chunks in lower-case}&\textbf{\# nodes}&\textbf{\# edges}\\
        \midrule
            KG-s1&10&no&10043&15470\\
            KG-s2&10&yes&10706&16179\\
            KG-s3&3&no&5306&7385\\
            KG-s4&3&yes&5172&6940\\
            KG-s5&5&no&6473&9300\\
            KG-s6&5&yes&6862&9521\\
        \bottomrule
    \end{tabular}
    \caption{Overview of the hyperparameter setups used to generate the different versions of the KG.}
    \label{tab:kg-setup}
\end{table}


\begin{figure}[!h]
    \centering
    \includegraphics[width=0.8\linewidth]{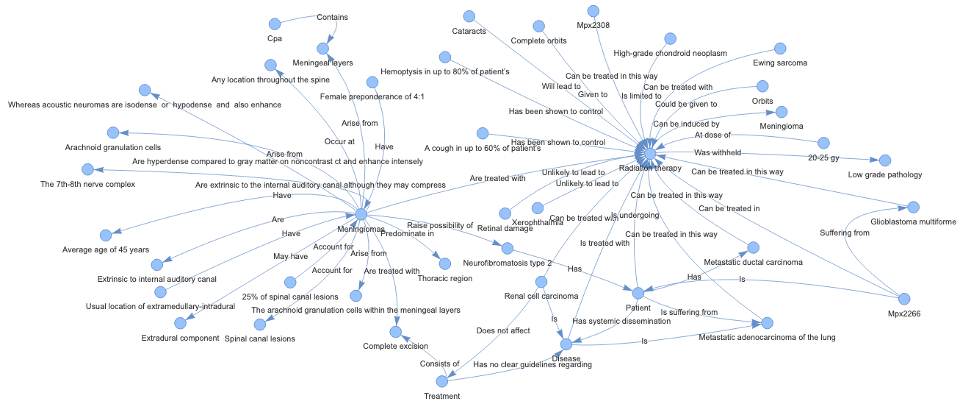}
    \caption{Overview of the subgraph generated from the node 'Meningiomas' in KG-s1.}
    \label{fig:subkg-overview}
\end{figure}

The entire process of creation and inference over the KGs was run on a cluster with 2 NVIDIA A100 64 GB GPUs.

\section{Results}\label{res}
In this section are reported the results obtained with the developed multimodal pipeline, where an end-to-end VLM is used for medical diagnosis support. The architecture of this system is depicted in Fig. \ref{fig:complete}, and couples the original DR-Minerva model with the KG built as reported in Section \ref{kg_definition}. The main contribution is to assess the effective usability of the presented data set for complex tasks such as the proposed multimodal pipeline.


\begin{figure}[!h]
    \centering
    \includegraphics[width=0.8\linewidth]{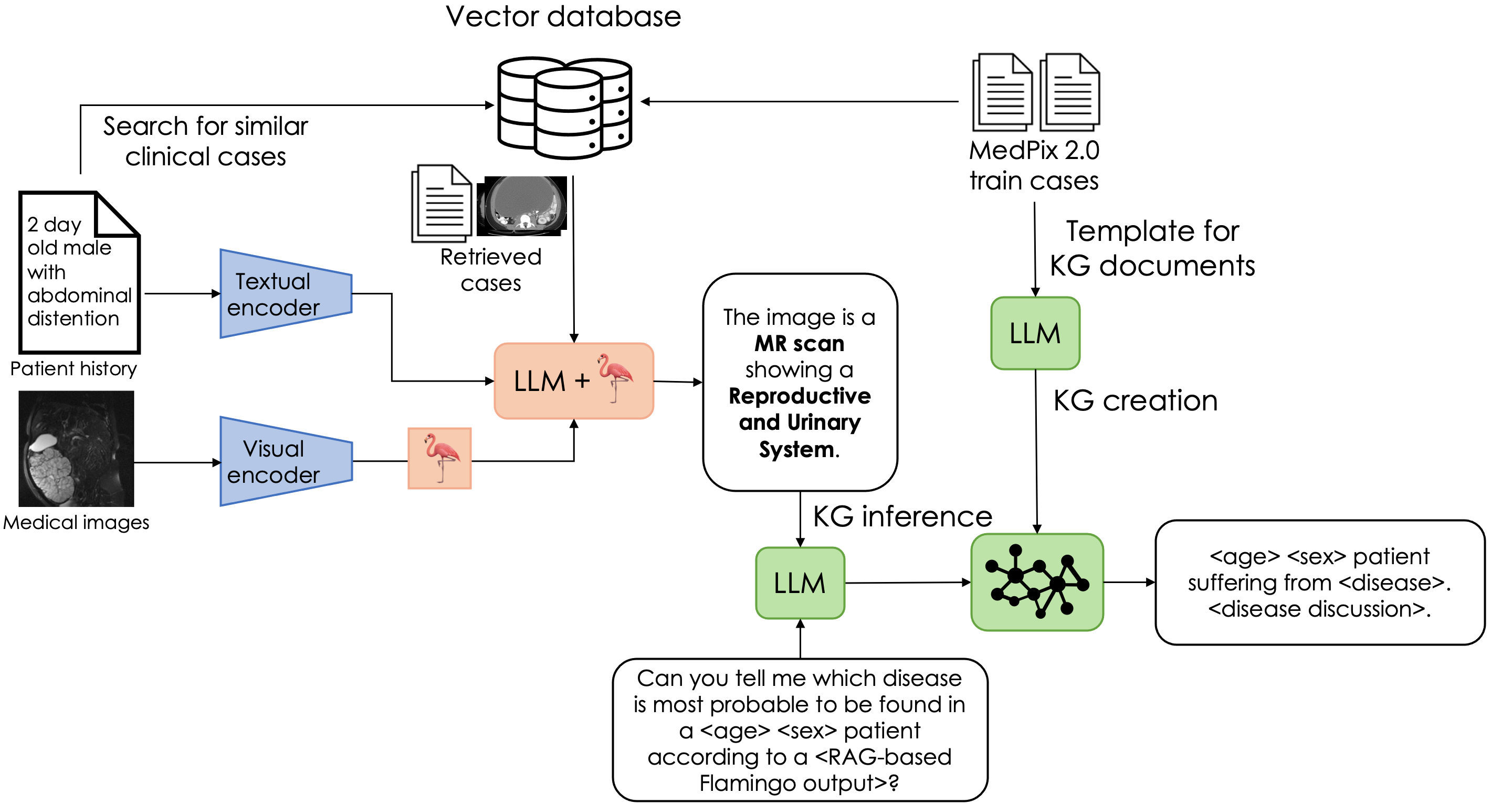}
    \caption{Overview of the generated architecture with RAG-based Flamingo and KG creation and inference.}
    \label{fig:complete}
\end{figure}

Performance evaluation was carried out by querying all the KGs with different hyperparameter choices, using prompts of increasing complexity. Specifically, three prompts were constructed, which were called \textit{simple}, \textit{mid-complex} and \textit{complex} respectively. All the prompts are reported as follows.

\begin{quote}
\noindent\fbox{%
    \parbox{0.92\textwidth}{%
Prompt \textit{simple}:\\
Can you tell me which diseases is most probable to be found in a patient having a \(\bigl \langle \text{DR-Minerva output} \bigr \rangle \)?
\\ \\
Prompt \textit{mid-complex}:\\
Can you tell me which diseases is most probable to be found in a \(\bigl \langle \text{age} \bigr \rangle \) \(\bigl \langle \text{sex} \bigr \rangle \) patient according to a \(\bigl \langle \text{DR-Minerva output}  \bigr \rangle \)?
\\  \\
Prompt \textit{complex}:\\
Can you tell me which diseases are the most probable to be found in a \(\bigl \langle \text{age} \bigr \rangle \) \(\bigl \langle \text{sex} \bigr \rangle \) patient according to a \(\bigl \langle \text{DR-Minerva output} \bigr \rangle \)? Consider also the following additional information about the patient. \(\bigl \langle \text{history} \bigr \rangle \).
    }%
}
\end{quote}

The three prompts use the information provided by DR-Minerva, which consists in the scanning modality and the body part shown in the input image. Prompts were designed to provide more and more information to the KG, as their complexity increases. The maximum complexity will be reached in the last prompt, where we want to emulate the query that a doctor might pose to a diagnostic support system. In this case, information concerning the patient's clinical \(\bigl \langle history \bigr \rangle \) is used in addition to the information extrapolated from the images supplied to the multimodal system.

The availability of a test set extrapolated from MedPix 2.0 allowed us to generate a golden answer according the following template.

\begin{quote}
\noindent\fbox{%
    \parbox{0.92\textwidth}{%
\centering
    \(\bigl \langle \text{age} \bigr \rangle \) \(\bigl \langle \text{sex} \bigr \rangle \) patient suffering from \(\bigl \langle \text{disease} \bigr \rangle \). \(\bigl \langle \text{disease discussion} \bigr \rangle \).
    }%
}
\end{quote}

Golden answers were used for the subsequent performance evaluation phase. Several metrics exist in the literature for automatic evaluation of LLM-generated text with respect to a set of reference or ``golden'' answers. These metrics score the degree of overlap between the reference answer $r = \{r_1,\ldots,r_m\}$ and the hypothesis $h 0 \{h_1,\ldots,h_n\}$ generated by the model. There are two approaches in this respect: character, word, or n-gram overlap and embedding similarity. BLEU \citep{bleu} and ROUGE \citep{rouge} belong to the first category. They are generally more tailored for evaluating Question-Answering and Machine Translation systems, for which a well-defined answer is known. Such measures require the generated text $h$ to adhere strictly to the golden answer $r$. Despite the validity of these metrics, are not meaningful for evaluation purposes. In this context, a semantically comparison is required for the proposed task, with other metrics.

In the process of generation clinical text to support a doctor in the diagnostic process, the same meaning can be expressed in a variety of forms that depend on many factors. Some of these factors are the available knowledge about the particular case when the diagnosis is formulated, the different clinical practices in writing medical records, and so on. As a consequence, despite the generated golden answers and the data set splits, we cannot expect the model to generate text exactly as the golden answer: multiple correct answers (as for the meaning) can be proposed with different linguistic nuances.



For these reasons, metrics based on embeddings similarity scores like METEOR \citep{meteor} and BERT score \citep{bertscore}, are preferred. In this case the generic $r_j$ and $h_i$ are embeddings and not words or n-grams. 
In particular, we decided to report in detail the BERT score in this paper and other metrics are reported in the supplementary material. BERT score is a neural framework to compare the embeddings of the tokens in $r$ and $h$ in terms of classical Precision, Recall, and F1 measure, which assess a compliant evaluation with the traditional metrics in AI field. In contrast to a traditional classification performance evaluation approach, Recall, Precision and BERT's F1 aim to evaluate the degree of overlap of the generated response with respect to the target response. In our specific case, this golden label is represented by the entire topic document containing the clinical information. Consequently, the BERT metrics are well suited to evaluating a specific task of the model, which is to provide a verbose response to the doctor that is as close as possible to the correct clinical terminology. Equation \ref{eq:bertscore} reports the BERTs score formulation.

\begin{equation}\label{eq:bertscore}
\begin{array}{lll}
&\text{BERT-P}&=\frac{1}{n}\sum_{i=1}^{n} \underset{r_j}{\max}\ \text{Sim}(h_i, r_j)\\ \\
&\text{BERT-R}&=\frac{1}{m}\sum_{j=1}^{m} \underset{h_i}{\max}\ \text{Sim}(r_j, h_i)\\ \\
&\text{BERT-F1}&=2 \cdot \frac{\text{BERT-P} \cdot \text{BERT-R}}{\text{BERT-P} + \text{BERT-R}}.
\end{array}
\end{equation}

$Sim(r_j, h_i)$ is the cosine similarity between the $j$-th reference token embedding and the $i$-th hypothesis token embedding, respectively. Table \ref{tab:tab_bert_score_sum} shows the results obtained using all the three prompts. We report the results with the other metrics (BLEU, ROUGE, METEOR) in the supplementary material. Here we only report BERT score since it is the most significant.

\begin{table}[!h]
    \centering
    \resizebox{\textwidth}{!}{\begin{tabular}{c|c|c|c|c|c|c|c|c|c|c}
    \toprule
    \multicolumn{2}{c|}{//}&\multicolumn{3}{c}{Simple}&\multicolumn{3}{|c|}{Mid}&\multicolumn{3}{|c}{Complex}\\
    \midrule
    KG-setup&Location&BERT-Precision&BERT-Recall&BERT-F1&BERT-Precision&BERT-Recall&BERT-F1&BERT-Precision&BERT-Recall&BERT-F1\\
    \midrule
    \multirow{5}{3em}{KG-s1}&Head&0.8027&0.7840&\textbf{0.7932}&0.7870&0.7848&0.7857&0.7971&0.7858&0.7913\\
    &Thorax&0.8003&0.7801&\textbf{0.79}&0.7853&0.7806&0.7827&0.7875&0.7822&0.7847\\
    &SaM&0.7888&0.7932&\textbf{0.791}&0.7964&0.7851&0.7905&0.7846&0.7885&0.7864\\
    &Abdomen&0.8008&0.7855&\textbf{0.793}&0.7859&0.7782&0.7818&0.7843&0.7817&0.7829\\
    &RUS&0.8050&0.7886&\textbf{0.7967}&0.7774&0.7787&0.778&0.7884&0.7766&0.7823\\
    \midrule
    \multirow{5}{3em}{KG-s2}&Head&0.8090&0.7900&\textbf{0.7994}&0.7985&0.7863&0.7921&0.7888&0.7802&0.7843\\
    &Thorax&0.8039&0.7851&\textbf{0.7943}&0.7981&0.7746&0.7856&0.7873&0.7821&0.7846\\
    &SaM&0.7981&0.7950&\textbf{0.7964}&0.7888&0.7908&0.7898&0.7949&0.7860&0.7903\\
    &Abdomen&0.7980&0.7845&\textbf{0.7912}&0.7882&0.7830&0.7853&0.7902&0.7863&0.7882\\
    &RUS&0.7971&0.7850&0.791&0.8011&0.7817&0.7912&0.8063&0.7890&\textbf{0.7975}\\
    \midrule
    \multirow{5}{3em}{KG-s3}&Head&0.7990&0.7855&0.7921&0.7991&0.7825&0.7905&0.8019&0.7872&\textbf{0.7944}\\
    &Thorax&0.7987&0.7840&\textbf{0.7912}&0.7885&0.7800&0.784&0.7858&0.7856&0.7855\\
    &SaM&0.8058&0.7958&\textbf{0.8007}&0.7820&0.7906&0.7862&0.7993&0.7902&0.7947\\
    &Abdomen&0.7960&0.7874&\textbf{0.7916}&0.7815&0.7742&0.7775&0.7844&0.7846&0.784\\
    &RUS&0.8021&0.7799&0.7908&0.7972&0.7781&0.7874&0.7985&0.7878&\textbf{0.7929}\\
    \midrule
    \multirow{5}{3em}{KG-s4}&Head&0.7989&0.7799&0.7891&0.7915&0.7829&0.787&0.7952&0.7843&\textbf{0.7895}\\
    &Thorax&0.8031&0.7784&\textbf{0.7905}&0.7869&0.7764&0.7815&0.7922&0.7822&0.7868\\
    &SaM&0.8023&0.7934&\textbf{0.7977}&0.7870&0.7874&0.7871&0.8006&0.7878&0.7939\\
    &Abdomen&0.7989&0.7769&0.7875&0.7975&0.7837&\textbf{0.7905}&0.7779&0.7765&0.7771\\
    &RUS&0.7979&0.7819&\textbf{0.7898}&0.7753&0.7798&0.7774&0.7937&0.7819&0.7877\\
    \midrule
    \multirow{5}{3em}{KG-s5}&Head&0.7999&0.7858&\textbf{0.7927}&0.7960&0.7876&0.7917&0.7945&0.7814&0.7877\\
    &Thorax&0.7960&0.7808&\textbf{0.7882}&0.7915&0.7769&0.784&0.7883&0.7827&0.7852\\
    &SaM&0.8037&0.7953&\textbf{0.7995}&0.7888&0.7932&0.7908&0.8049&0.7925&0.7985\\
    &Abdomen&0.8012&0.7772&0.7889&0.7840&0.7795&0.7816&0.8028&0.7820&\textbf{0.7921}\\
    &RUS&0.7861&0.7860&0.786&0.7876&0.7871&0.7873&0.8025&0.7898&\textbf{0.796}\\
    \midrule
    \multirow{5}{3em}{KG-s6}&Head&0.7910&0.7864&\textbf{0.7885}&0.7923&0.7801&0.786&0.7931&0.7799&0.7862\\
    &Thorax&0.7859&0.7812&0.7835&0.7720&0.7754&0.7735&0.7879&0.7808&\textbf{0.7842}\\
    &SaM&0.7832&0.7893&0.7861&0.7957&0.7893&0.7924&0.8106&0.7918&\textbf{0.8009}\\
    &Abdomen&0.7807&0.7816&0.7811&0.7958&0.7875&\textbf{0.7916}&0.7886&0.7801&0.7841\\
    &RUS&0.7973&0.7826&0.7898&0.7849&0.7842&0.7845&0.7994&0.7886&\textbf{0.7938}\\
    \bottomrule
    \end{tabular}}
    \caption{Summary of BERT metrics calculated on all KG setups for each of the proposed prompts. Bold values indicate the best score for each KG and body part respectively}
    \label{tab:tab_bert_score_sum}
\end{table}

The results show that the prompt \textit{simple} achieves best performance for the models using KG-s1 and KG-s2, that are the KGs with the highest number of relations per chunk. The prompt \textit{complex} performs best for the model using KG-s6, and it is close to the prompt \textit{simple} for the models using KG-s3. No particular prompt prevails over the others in the models using KG-s4 and KG-s5.

Given the structure of the KGs, one can argue that the \textit{simple} prompt does not constrain the model too much in the generation phase, and allows for the construction of verbose responses whose context is controlled by the high number of connected document chunks in the KG. On the other end, when the KG connects less document chunks, the best performance of the prompt \textit{complex} in terms of BERT-F1 depends on the increased generation precision induced by the insertion of the patient \(\bigl \langle \text{history} \bigr \rangle \).

At inference time, Llama 3.1 8B Instruct is invoked for answer generation by retrieving relevant information from the KGs, leveraging the given question. We used a low temperature of $0.00001$, thus the generative capabilities of the model are limited, and then, it is forced to produce an answer as much close as possible to the retrieved information. Moreover, to prevent the model in generating loop, the \texttt{no\_repeat\_ngram\_size} parameter was set to $2$. The query strategy over the KG was done using the \texttt{tree\_summarize} response mode and a value equal to $5$ as for \texttt{similarity\_top\_k} parameter, that is the default parameters for the retriever. We report four end-to-end inferences using the better prompts and KGs, namely simple prompt and KG-s1 (Fig \ref{fig:simple-s1}) and complex prompt and KG-s6 \ref{fig:complex-s6}). Both the image, the input prompt to the model, the generated answer and the BERT-F1 score are reported.

\begin{figure}[!h]
    \centering
    \includegraphics[width=\linewidth]{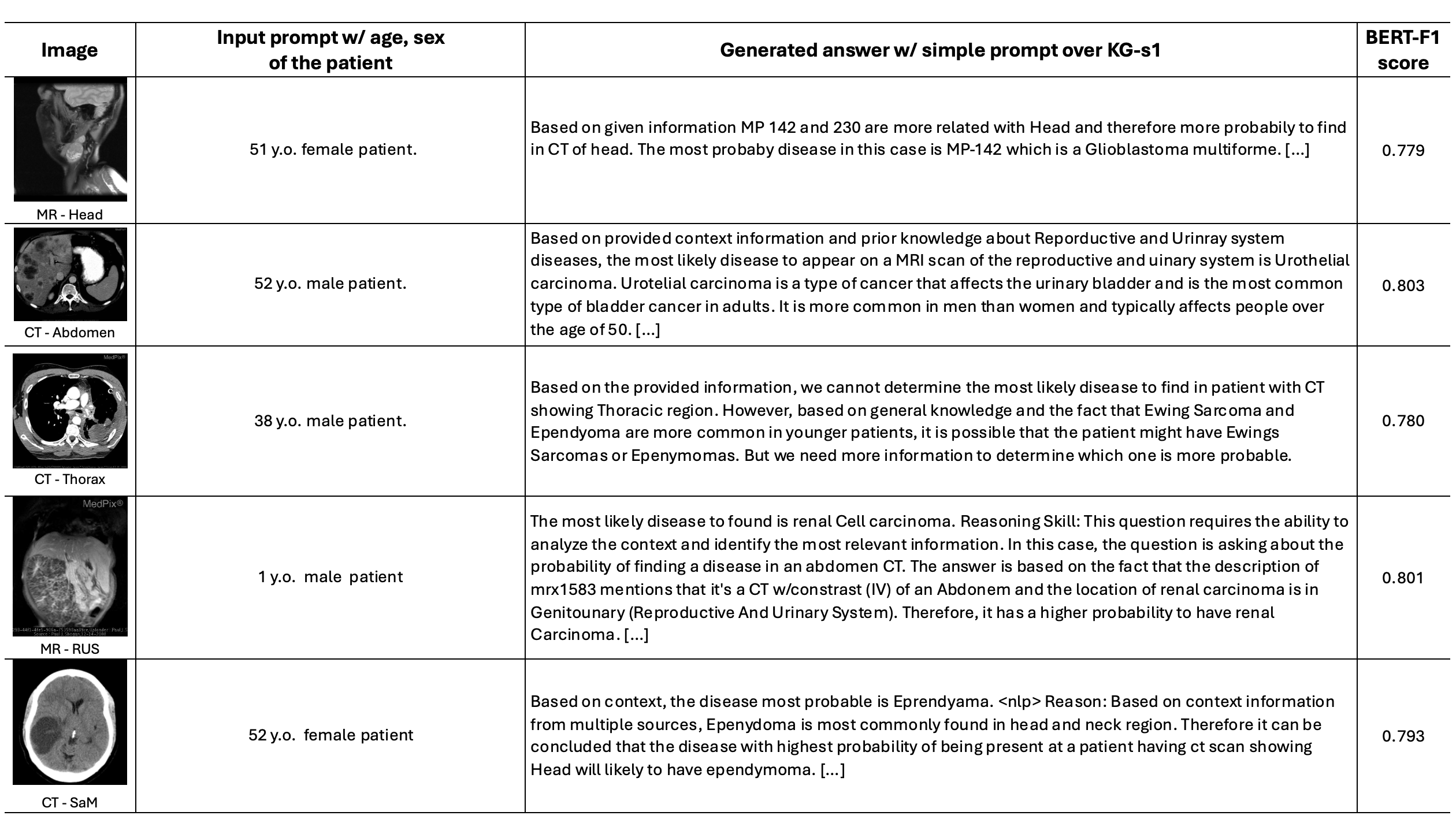}
    \caption{One example for each body part of end-to-end queries to the entire system leveraging KG-s1 and simple prompt, along with inputs and the final generated answer.}
    \label{fig:simple-s1}
\end{figure}

\begin{figure}[!h]
    \centering
    \includegraphics[width=\linewidth]{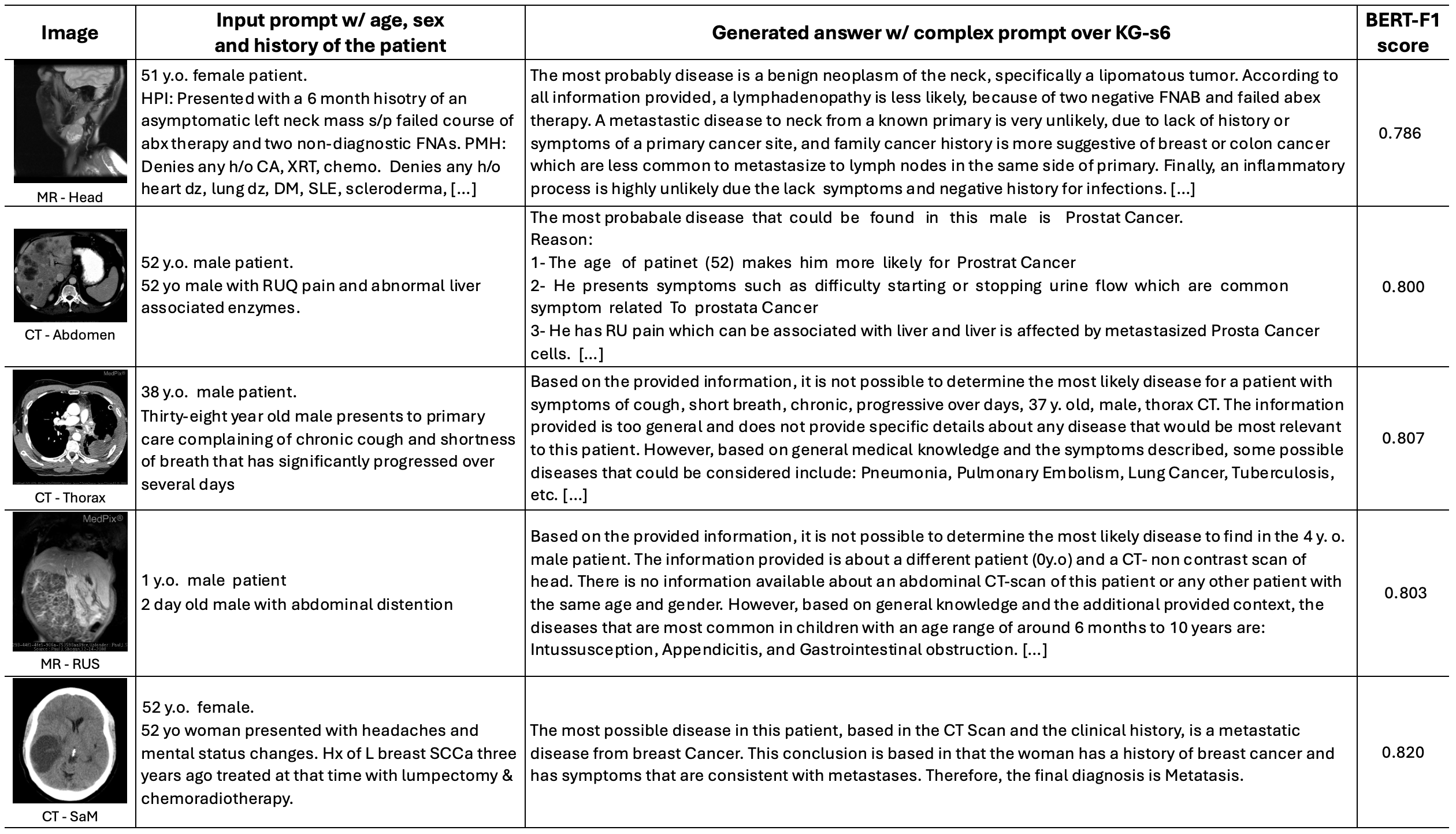}
    \caption{One example for each body part of end-to-end queries to the entire system leveraging KG-s6 and complex prompt, along with inputs and the final generated answer.}
    \label{fig:complex-s6}
\end{figure}

\section{Limitations}\label{limitations}

MedPix 2.0 is proposed as a multimodal data set for training deep neural architectures. To demonstrate the robustness and ease of implementing the data set, a model was proposed that combined a RAG-based Flamingo, Dr-Minerva, (created by the authors in a previous work) in combination with Llama 3.1 for the generation of verbose responses. As can be seen from the results obtained and evaluated in terms of BERT Precision, Recall and F1, the result is robust enough to allow an initial experimental version of a multimodal model to achieve good performance in generating a coherent verbose response. Despite the promising results, the construction of a reliable model is not the main topic of the proposed work, which is why validation by doctors has not been carried out, which is a fundamental step in proposing a model to support diagnosis. Despite the simplicity with which MedPix 2.0 can be queried and used in its Image-only, Text-only and Multimodal versions, its integration with other data sets can improve its structure, making it more complete and maximising its effectiveness in AI training tasks for diagnostic support.


\section{Conclusions and Future work}\label{future_work}
In this work we presented MedPix 2.0, a multimodal data set of clinical reports, CT and MR scans. We devised a semi-automated pipeline to download and curate the images in the original data sets, while structuring the textual information as a set of JSON document collections which had been used to build a proper MongoDB instance. The NoSQL version of the data set can be accessed and queried with a usable GUI that has been developed purposely. Using the GUI one can browse the data set in the same manner as in the original data set, and can download the structured output of the query that is suitable for training AI models. In fact, 
MedPix\textsuperscript{\textregistered} and its MongoDB interface, represent in our view a relevant starting point for the development of AI multimodal models in the medical domain. such as Information Extraction systems tailored for clinical reports, automated analysis of the medical images, or Generative AI models for clinical report generation as part of a Medical Decision Support System. All these systems can rely on MedPix 2.0 as a structured source of data containing both clinical cases and medical explanations about the disease under investigation.

To demonstrate this point, we recalled the architecture of DR-Minerva, a RAG-based Flamingo architecture already developed by the authors, and presented a brand new VLM architecture that couples DR-Minerva with a KG created automatically relying on Llama 3.1 8B Instruct. Different KG structures were tested, starting from the MedPix 2.0 test set along with prompts of increasing complexity as regards the context provided to the model. The results we obtained are very satisfactory but we are actively working to improve the model generation ability by means of a structural and semi-automatic organization of the KG, relying on external medical ontologies.

The scalability of the developed MongoDB database, makes it suitable for future extensions with the possibility to add brand new clinical cases or existing cases from data set mentioned in Section \ref{sota}, that have to be compliant with privacy regulations and follow the required information structure. Moreover, the inherent distributed nature of MongoDB allows for creating huge databases across different wards where the data owned by a single institution do not need to be explicitly moved out of the hospital thus violating privacy regulations. The structure of MedPix 2.0 could also serve as a guide to develop suitable connectors to share allowed data in the EHDS.

New cases can also be easily added to the KG, thus expanding the clinical knowledge base used by the system in the inference phase. Further improvement is being done on the GUI, providing the user with advanced data visualization tools like the possibility to interactively compare similar cases, thus helping physicians during the diagnostic phase.

\subsection*{Supplementary information}
Tables with the results obtained from the experiments for the clinical cases for the three prompts and for all KG setups are given in the supplementary file.

\subsection*{Acknowledgements}
This work is supported by the project B73C22000810001, project code ECS\_00000022, “SAMOTHRACE” (Sicilian MicronanoTech Research And Innovation Center). \\Models are built on the Leonardo supercomputer with the support of CINECA-Italian Super Computing Resource Allocation, class C project IscrC\_DOC-VLM (HP10C64J82).

%
%
%
\bibliographystyle{plain}
\bibliography{myref}

\begin{thebibliography}{10}

\bibitem{flamingo}
Jean-Baptiste Alayrac, Jeff Donahue, Pauline Luc, Antoine Miech, Iain Barr, Yana Hasson, Karel Lenc, Arthur Mensch, Katie Millican, Malcolm Reynolds, Roman Ring, Eliza Rutherford, Serkan Cabi, Tengda Han, Zhitao Gong, Sina Samangooei, Marianne Monteiro, Jacob Menick, Sebastian Borgeaud, Andrew Brock, Aida Nematzadeh, Sahand Sharifzadeh, Mikolaj Binkowski, Ricardo Barreira, Oriol Vinyals, Andrew Zisserman, and Karen Simonyan.
\newblock {Flamingo: a Visual Language Model for Few-Shot Learning}, 2022.

\bibitem{awadalla2023openflamingo}
Anas Awadalla, Irena Gao, Josh Gardner, Jack Hessel, Yusuf Hanafy, Wanrong Zhu, Kalyani Marathe, Yonatan Bitton, Samir Gadre, Shiori Sagawa, Jenia Jitsev, Simon Kornblith, Pang~Wei Koh, Gabriel Ilharco, Mitchell Wortsman, and Ludwig Schmidt.
\newblock {OpenFlamingo: An Open-Source Framework for Training Large Autoregressive Vision-Language Models}.
\newblock {\em arXiv preprint arXiv:2308.01390}, 2023.

\bibitem{meteor}
Satanjeev Banerjee and Alon Lavie.
\newblock {METEOR: An automatic metric for MT evaluation with improved correlation with human judgments}.
\newblock In {\em Proceedings of the acl workshop on intrinsic and extrinsic evaluation measures for machine translation and/or summarization}, pages 65--72, 2005.

\bibitem{gpt3}
Tom Brown, Benjamin Mann, Nick Ryder, Melanie Subbiah, Jared~D Kaplan, Prafulla Dhariwal, Arvind Neelakantan, Pranav Shyam, Girish Sastry, Amanda Askell, et~al.
\newblock {Language Models are Few-Shot Learners}.
\newblock {\em Advances in neural information processing systems}, 2020.

\bibitem{chaudhry2022unitochest}
Hafiza Ayesha~Hoor Chaudhry, Riccardo Renzulli, Daniele Perlo, Francesca Santinelli, Stefano Tibaldi, Carmen Cristiano, Marco Grosso, Giorgio Limerutti, Attilio Fiandrotti, Marco Grangetto, et~al.
\newblock {Unitochest: A lung image dataset for segmentation of cancerous nodules on ct scans}.
\newblock In {\em International Conference on Image Analysis and Processing}, pages 185--196. Springer, 2022.

\bibitem{Chevrier_2019}
Raphaël Chevrier, Vasiliki Foufi, Christophe Gaudet-Blavignac, Arnaud Robert, and Christian Lovis.
\newblock Use and understanding of anonymization and de-identification in the biomedical literature: Scoping review.
\newblock {\em Journal of Medical Internet Research}, 21(5):e13484, May 2019.

\bibitem{demner2016preparing}
Dina Demner-Fushman, Marc~D Kohli, Marc~B Rosenman, Sonya~E Shooshan, Laritza Rodriguez, Sameer Antani, George~R Thoma, and Clement~J McDonald.
\newblock {Preparing a collection of radiology examinations for distribution and retrieval}.
\newblock {\em Journal of the American Medical Informatics Association}, 23(2):304--310, 2016.

\bibitem{faiss}
Matthijs Douze, Alexandr Guzhva, Chengqi Deng, Jeff Johnson, Gergely Szilvasy, Pierre-Emmanuel Mazaré, Maria Lomeli, Lucas Hosseini, and Hervé Jégou.
\newblock {The Faiss library}, 2024.

\bibitem{El_Emam_2015}
Khaled El~Emam, Sam Rodgers, and Bradley Malin.
\newblock Anonymising and sharing individual patient data.
\newblock {\em The BMJ}, 350:h1139, March 2015.

\bibitem{unitobrain}
Umberto Gava, Federico D'Agata, Edwin Bennink, Enzo Tartaglione, Daniele Perlo, Annamaria Vernone, Francesca Bertolino, Eleonora Ficiarà, Alessandro Cicerale, Fabrizio Pizzagalli, Caterina Guiot, Marco Grangetto, and Mauro Bergui.
\newblock {UniTOBrain}, 2021.

\bibitem{medchat}
Saba Ghanbari~Haez, Marina Segala, Patrizio Bellan, Simone Magnolini, Leonardo Sanna, Monica Consolandi, and Mauro Dragoni.
\newblock {A Retrieval-Augmented Generation Strategy to Enhance Medical Chatbot Reliability}.
\newblock In {\em Artificial Intelligence in Medicine}, 2024.

\bibitem{OPTIMAM}
Mark~D. Halling-Brown, Lucy~M. Warren, Dominic Ward, Emma Lewis, Alistair Mackenzie, Matthew~G. Wallis, Louise~S. Wilkinson, Rosalind~M. Given-Wilson, Rita McAvinchey, and Kenneth~C. Young.
\newblock {OPTIMAM Mammography Image Database: A Large-Scale Resource of Mammography Images and Clinical Data}.
\newblock {\em Radiology: Artificial Intelligence}, 3(1):e200103, 2021.
\newblock PMID: 33937853.

\bibitem{he2020pathvqa}
Xuehai He, Yichen Zhang, Luntian Mou, Eric Xing, and Pengtao Xie.
\newblock Pathvqa: 30000+ questions for medical visual question answering.
\newblock {\em arXiv preprint arXiv:2003.10286}, 2020.

\bibitem{huang2023inspect}
Shih-Cheng Huang, Zepeng Huo, Ethan Steinberg, Chia-Chun Chiang, Matthew~P Lungren, Curtis~P Langlotz, Serena Yeung, Nigam~H Shah, and Jason~A Fries.
\newblock {INSPECT: a multimodal dataset for pulmonary embolism diagnosis and prognosis}.
\newblock {\em arXiv preprint arXiv:2311.10798}, 2023.

\bibitem{adni}
Clifford~R. Jack~Jr., Matt~A. Bernstein, Nick~C. Fox, Paul Thompson, Gene Alexander, Danielle Harvey, Bret Borowski, Paula~J. Britson, Jennifer L.~Whitwell, Chadwick Ward, Anders~M. Dale, Joel~P. Felmlee, Jeffrey~L. Gunter, Derek~L.G. Hill, Ron Killiany, Norbert Schuff, Sabrina Fox-Bosetti, Chen Lin, Colin Studholme, Charles~S. DeCarli, Gunnar Krueger, Heidi~A. Ward, Gregory~J. Metzger, Katherine~T. Scott, Richard Mallozzi, Daniel Blezek, Joshua Levy, Josef~P. Debbins, Adam~S. Fleisher, Marilyn Albert, Robert Green, George Bartzokis, Gary Glover, John Mugler, and Michael~W. Weiner.
\newblock {The Alzheimer's disease neuroimaging initiative (ADNI): MRI methods}.
\newblock {\em Journal of Magnetic Resonance Imaging}, 27(4):685--691, 2008.

\bibitem{jiang2023mistral7b}
Albert~Q. Jiang, Alexandre Sablayrolles, Arthur Mensch, Chris Bamford, Devendra~Singh Chaplot, Diego de~las Casas, Florian Bressand, Gianna Lengyel, Guillaume Lample, Lucile Saulnier, Lélio~Renard Lavaud, Marie-Anne Lachaux, Pierre Stock, Teven~Le Scao, Thibaut Lavril, Thomas Wang, Timothée Lacroix, and William~El Sayed.
\newblock {Mistral 7B}, 2023.

\bibitem{Jiang_Fang_2024}
Xinke Jiang, Yue Fang, Rihong Qiu, Haoyu Zhang, Yongxin Xu, Hao Chen, Wentao Zhang, Ruizhe Zhang, Yuchen Fang, Xu~Chu, Junfeng Zhao, and Yasha Wang.
\newblock Tc-rag:turing-complete rag’s case study on medical llm systems.
\newblock (arXiv:2408.09199), August 2024.

\bibitem{johnson2019mimic}
Alistair~EW Johnson, Tom~J Pollard, Seth~J Berkowitz, Nathaniel~R Greenbaum, Matthew~P Lungren, Chih-ying Deng, Roger~G Mark, and Steven Horng.
\newblock {MIMIC-CXR, a de-identified publicly available database of chest radiographs with free-text reports}.
\newblock {\em Scientific data}, 6(1):317, 2019.

\bibitem{embedMistral}
Kim Junseong, Lee Seolhwa, Kwon Jihoon, Gu~Sangmo, Kim Yejin, Cho Minkyung, Sohn Jy-yong, and Choi Chanyeol.
\newblock {Linq-Embed-Mistral:Elevating Text Retrieval with Improved GPT Data Through Task-Specific Control and Quality Refinement}.
\newblock Linq AI Research Blog, 2024.

\bibitem{oasis3}
Pamela~J. LaMontagne, Tammie~LS. Benzinger, John~C. Morris, Sarah Keefe, Russ Hornbeck, Chengjie Xiong, Elizabeth Grant, Jason Hassenstab, Krista Moulder, Andrei~G. Vlassenko, Marcus~E. Raichle, Carlos Cruchaga, and Daniel Marcus.
\newblock {OASIS-3: Longitudinal Neuroimaging, Clinical, and Cognitive Dataset for Normal Aging and Alzheimer Disease}.
\newblock {\em medRxiv}, 2019.

\bibitem{lau2018dataset}
Jason~J Lau, Soumya Gayen, Asma Ben~Abacha, and Dina Demner-Fushman.
\newblock {A dataset of clinically generated visual questions and answers about radiology images}.
\newblock {\em Scientific data}, 5(1):1--10, 2018.

\bibitem{Lee_2024}
Jungwon Lee, Seungjun Ahn, Daeho Kim, and Dongkyun Kim.
\newblock Performance comparison of retrieval-augmented generation and fine-tuned large language models for construction safety management knowledge retrieval.
\newblock {\em Automation in Construction}, 168:105846, December 2024.

\bibitem{rag}
Patrick Lewis, Ethan Perez, Aleksandra Piktus, Fabio Petroni, Vladimir Karpukhin, Naman Goyal, Heinrich K{\"u}ttler, Mike Lewis, Wen-tau Yih, Tim Rockt{\"a}schel, et~al.
\newblock {Retrieval-Augmented Generation for Knowledge-Intensive NLP Tasks}.
\newblock {\em Advances in Neural Information Processing Systems}, 2020.

\bibitem{rouge}
Chin-Yew Lin.
\newblock {Rouge: A package for automatic evaluation of summaries}.
\newblock In {\em Text summarization branches out}, pages 74--81, 2004.

\bibitem{lin2014mscoco}
Tsung-Yi Lin, Michael Maire, Serge Belongie, James Hays, Pietro Perona, Deva Ramanan, Piotr Doll{\'a}r, and C~Lawrence Zitnick.
\newblock {Microsoft coco: Common objects in context}.
\newblock In {\em Computer Vision--ECCV 2014: 13th European Conference, Zurich, Switzerland, September 6-12, 2014, Proceedings, Part V 13}, pages 740--755. Springer, 2014.

\bibitem{lin2023pmc}
Weixiong Lin, Ziheng Zhao, Xiaoman Zhang, Chaoyi Wu, Ya~Zhang, Yanfeng Wang, and Weidi Xie.
\newblock {Pmc-clip: Contrastive language-image pre-training using biomedical documents}.
\newblock In {\em International Conference on Medical Image Computing and Computer-Assisted Intervention}, pages 525--536. Springer, 2023.

\bibitem{liu2021slake}
Bo~Liu, Li-Ming Zhan, Li~Xu, Lin Ma, Yan Yang, and Xiao-Ming Wu.
\newblock {Slake: A semantically-labeled knowledge-enhanced dataset for medical visual question answering}.
\newblock In {\em 2021 IEEE 18th International Symposium on Biomedical Imaging (ISBI)}, pages 1650--1654. IEEE, 2021.

\bibitem{Liu_LlamaIndex_2022}
Jerry Liu.
\newblock {LlamaIndex}, 11 2022.

\bibitem{llama3}
AI~@~Meta Llama~Team.
\newblock {The Llama 3 Herd of Models}, 2024.

\bibitem{Magnini2020TheEP}
Bernardo Magnini, Bego{\~n}a Altuna, Alberto Lavelli, Manuela Speranza, and Roberto Zanoli.
\newblock {The E3C Project: Collection and Annotation of a Multilingual Corpus of Clinical Cases}.
\newblock {\em Proceedings of the Seventh Italian Conference on Computational Linguistics CLiC-it 2020}, 2020.

\bibitem{agmodels}
Grégoire Mialon, Roberto Dessì, Maria Lomeli, Christoforos Nalmpantis, Ram Pasunuru, Roberta Raileanu, Baptiste Rozière, Timo Schick, Jane Dwivedi-Yu, Asli Celikyilmaz, Edouard Grave, Yann LeCun, and Thomas Scialom.
\newblock {Augmented Language Models: a Survey}, 2023.

\bibitem{muennighoff2022mteb}
Niklas Muennighoff, Nouamane Tazi, Lo{\"\i}c Magne, and Nils Reimers.
\newblock {MTEB: Massive Text Embedding Benchmark}.
\newblock {\em arXiv preprint arXiv:2210.07316}, 2022.

\bibitem{minerva}
Riccardo Orlando, Luca Moroni, Pere-Llu{\'\i}s~H. Cabot, Edoardo Barba, Simone Conia, Sergio Orlandini, Giuseppe Fiameni, and Roberto Navigli.
\newblock {Minerva LLMs: The First Family of Large Language Models Trained from Scratch on Italian Data}.
\newblock {\em Proceedings of the 10th Italian Conference on Computational Linguistics (CLiC-it 2024)}, 2024.

\bibitem{bleu}
Kishore Papineni, Salim Roukos, Todd Ward, and Wei-Jing Zhu.
\newblock {BLEU: a method for automatic evaluation of machine translation}.
\newblock In {\em Proceedings of the 40th annual meeting of the Association for Computational Linguistics}, pages 311--318, 2002.

\bibitem{pelka2018radiology}
Obioma Pelka, Sven Koitka, Johannes R{\"u}ckert, Felix Nensa, and Christoph~M Friedrich.
\newblock {Radiology objects in context (roco): a multimodal image dataset}.
\newblock In {\em Intravascular Imaging and Computer Assisted Stenting and Large-Scale Annotation of Biomedical Data and Expert Label Synthesis: 7th Joint International Workshop, CVII-STENT 2018 and Third International Workshop, LABELS 2018, Held in Conjunction with MICCAI 2018, Granada, Spain, September 16, 2018, Proceedings 3}, pages 180--189. Springer, 2018.

\bibitem{penedo_2024}
Andrés~Chomczyk Penedo.
\newblock {The Regulation of Data Spaces under the EU Data Strategy: Towards the ‘Act-ification’ of the Fifth European Freedom for Data?}
\newblock {\em European Journal of Law and Technology}, 15(1), May 2024.

\bibitem{peng2023checkfactstryagain}
Baolin Peng, Michel Galley, Pengcheng He, Hao Cheng, Yujia Xie, Yu~Hu, Qiuyuan Huang, Lars Liden, Zhou Yu, Weizhu Chen, and Jianfeng Gao.
\newblock {Check Your Facts and Try Again: Improving Large Language Models with External Knowledge and Automated Feedback}, 2023.

\bibitem{clip}
Alec Radford, Jong~Wook Kim, Chris Hallacy, Aditya Ramesh, Gabriel Goh, Sandhini Agarwal, Girish Sastry, Amanda Askell, Pamela Mishkin, Jack Clark, Gretchen Krueger, and Ilya Sutskever.
\newblock {Learning Transferable Visual Models From Natural Language Supervision}, 2021.

\bibitem{dpo}
Rafael Rafailov, Archit Sharma, Eric Mitchell, Stefano Ermon, Christopher~D. Manning, and Chelsea Finn.
\newblock {Direct Preference Optimization: Your Language Model is Secretly a Reward Model}, 2024.

\bibitem{Salemi_Zamani_2024}
Alireza Salemi and Hamed Zamani.
\newblock Comparing retrieval-augmentation and parameter-efficient fine-tuning for privacy-preserving personalization of large language models.
\newblock (arXiv:2409.09510), September 2024.
\newblock arXiv:2409.09510 [cs].

\bibitem{schulz-etal-2020-named}
Sarah Schulz, Jurica {\v{S}}eva, Samuel Rodriguez, Malte Ostendorff, and Georg Rehm.
\newblock {Named Entities in Medical Case Reports: Corpus and Experiments}.
\newblock In {\em Proceedings of the Twelfth Language Resources and Evaluation Conference}, pages 4495--4500, Marseille, France, May 2020. European Language Resources Association.

\bibitem{drMinerva}
Irene Siragusa, Salvatore Contino, and Roberto Pirrone.
\newblock {DR-Minerva: a Multimodal Language Model based on Minerva for Diagnostic Information Retrieval}.
\newblock In {\em AIxIA 2024 -- Advances in Artificial Intelligence}, pages 288--300, Cham, 2025. Springer Nature Switzerland.

\bibitem{subramanian2020medicat}
Sanjay Subramanian, Lucy~Lu Wang, Sachin Mehta, Ben Bogin, Madeleine van Zuylen, Sravanthi Parasa, Sameer Singh, Matt Gardner, and Hannaneh Hajishirzi.
\newblock {Medicat: A dataset of medical images, captions, and textual references}.
\newblock {\em arXiv preprint arXiv:2010.06000}, 2020.

\bibitem{terzis_2023}
Petros Terzis and (Enrique) OE~Santamaria Echeverria.
\newblock {Interoperability and governance in the European Health Data Space regulation}.
\newblock {\em Medical Law International}, April 2023.

\bibitem{attention2017}
Ashish Vaswani, Noam Shazeer, Niki Parmar, Jakob Uszkoreit, Llion Jones, Aidan~N. Gomez, \L{}ukasz Kaiser, and Illia Polosukhin.
\newblock Attention is all you need.
\newblock In {\em Proceedings of the 31st International Conference on Neural Information Processing Systems}, NIPS'17, page 6000–6010, Red Hook, NY, USA, 2017. Curran Associates Inc.

\bibitem{vidivelli2024efficiency}
S~Vidivelli, Manikandan Ramachandran, and A~Dharunbalaji.
\newblock {Efficiency-Driven Custom Chatbot Development: Unleashing LangChain, RAG, and Performance-Optimized LLM Fusion}.
\newblock {\em Computers, Materials \& Continua}, 2024.

\bibitem{wang2024potential}
Calvin Wang, Joshua Ong, Chara Wang, Hannah Ong, Rebekah Cheng, and Dennis Ong.
\newblock {Potential for GPT technology to optimize future clinical decision-making using retrieval-augmented generation}.
\newblock {\em Annals of Biomedical Engineering}, 2024.

\bibitem{wu2024medicalgraphragsafe}
Junde Wu, Jiayuan Zhu, Yunli Qi, Jingkun Chen, Min Xu, Filippo Menolascina, and Vicente Grau.
\newblock Medical graph rag: Towards safe medical large language model via graph retrieval-augmented generation, 2024.

\bibitem{bertscore}
Tianyi Zhang, Varsha Kishore*, Felix Wu, Kilian~Q. Weinberger, and Yoav Artzi.
\newblock {BERTScore: Evaluating Text Generation with BERT}.
\newblock In {\em International Conference on Learning Representations}, 2020.

\end{thebibliography}

\end{document}